\documentclass{aa}

\usepackage{float}
\usepackage{subcaption}
\usepackage{graphicx}
\graphicspath{{Figures/}}
\usepackage[svgnames]{xcolor}
\usepackage{natbib}
\usepackage{amsmath}
\usepackage[varg]{txfonts}
\usepackage{caption}
\usepackage[pdftex,colorlinks,citecolor=MidnightBlue,linkcolor=MidnightBlue]{hyperref}
\usepackage{bm} %Pour avoir la police math en italique ET en gras, sinon utiliser \mathbf{•} mais ca enleve l'italique des lettres

\renewcommand{\d}{\, \mathrm{d}}

\newcommand{\Msun}{\ensuremath{\text{M}_\odot}}

\newcommand{\ramses}{\ensuremath{\mathcal{R}}\textsc{amses}}

\usepackage{siunitx}
\DeclareSIUnit\parsec{pc}
\DeclareSIUnit\lightyear{ly}
\DeclareSIUnit\year{yr}
\DeclareSIUnit{\Msun}{M_\odot}

\bibpunct{(}{)}{;}{a}{}{,} % to follow the A&A style
\begin{document}

  %\title{Origin of angular momentum in disks and dense cores.}
  %\title{Formation of protostellar disk by non-axisymmetrical gravitationnal collapse.}
  %\title{Formation of protostellar disk by gravitational collapse of a non rotating, non axisymmetrical cloud.}
  %\title{Formation of planet-forming disk by gravitational collapse of a non-rotating, non-axisymmetrical cloud.}
  %\title{Formation of protostellar disk by gravitational collapse of a non-rotating, non-axisymmetrical cloud.}
  %\title{An alternative model for protostellar disk formation: \\ gravitational collapse of a non-rotating, non-axisymmetrical cloud.}
  \title{Protostellar disk formation by a non-rotating, non-axisymmetric collapsing cloud: model and comparison with observations.}
  
  \titlerunning{Disk formation in collapsing non-axisymmetrical cores}
  
  \author{Antoine Verliat\inst{\ref{inst1}} 
    \and Patrick Hennebelle\inst{\ref{inst1},\ref{inst2}}
    \and Anaëlle J. Maury \inst{\ref{inst1},\ref{inst3}}
    \and Mathilde Gaudel \inst{\ref{inst1}}}

  \institute{Laboratoire AIM, Paris-Saclay, CEA Saclay/IRFU/DAp -- CNRS --
    Universit\'e Paris Diderot, 91191 Gif-sur-Yvette Cedex, France
    %Laboratoire AIM, CEA Saclay/IRFU/DAp, CNRS, Université Paris-Saclay, Université Paris Diderot, F-91191 Gif-sur-Yvette, France
    \label{inst1}
    \and
    LERMA (UMR CNRS 8112), Ecole Normale Sup\'erieure, 75231 Paris Cedex, France
    \label{inst2}
    \and 
	Harvard-Smithsonian Center for Astrophysics, 60 Garden street, Cambridge, MA 02138, USA
	\label{inst3}
    }

  \date{Received ---; accepted ---}

    \abstract
    {%Context:
    Planet-forming disks are fundamental objects thought to be inherited from large scale rotation, through the conservation of angular momentum during the collapse of a prestellar dense core.}
    {%Aims:
    We investigate the possibility for a protostellar disk to be formed from a motionless dense core which contains non-axisymmetric density fluctuations. The rotation is thus generated locally by the asymmetry of the collapse.}
    {%Methods:
      We study the evolution of the angular momentum in a non-axisymmetric collapse of a dense core from an analytical point of view. To test the theory, we perform three-dimensional simulations of a collapsing prestellar dense core using adaptative mesh refinement. We start from a non-axisymmetrical situation, considering a dense core with random density perturbations that follow a turbulence spectrum. We analyse the emerging disk comparing the angular momentum it contains with the one expected from our analytic development. We study the velocity gradients at different scales in the simulation as it is done with observations.}
    {%Results:
     %We show that the angular momentum with respect to a star that is not located at the center of mass of the cloud is not conserved, due to inertial forces. The simulations of a non axisymmetrical collapse show that the formation of a disk is compatible with this analysis. The analysis of orientation of velocity gradients at different scales shows that velocity gradients at small and large scales are not expected to be perfectly aligned. Velocity gradients inversions are also observed. This is a very distinct feature in comparison to model where rotation is inherited from large scales. The value of the apparent specific angular momentum is roughly constant over the different scales and match with observational results that are usually interpreted as angular momentum conservation.
     We show that the angular momentum in the frame of a stellar object which is not located at the center of mass of the core is not conserved, due to inertial forces. Our simulations of such non-axisymmetrical collapse quickly produce accretion disks at the small scales in the core. The analysis of the kinematics at different scales in the simulated core reveals projected velocity gradients of amplitudes similar to the ones observed in protostellar cores, and which directions vary, sometimes even reversing when small and large scales are compared. These complex kinematics patterns appear in recent observations, and could be a discriminating feature with models where rotation is inherited from large scales. Our results from simulations without initial rotation are more consistent with these recent observations than when solid-body rotation is initially imprinted. Lastly, we show
     that the disks formed in this scenario of non-axisymmetrical gravitational collapse grow to reach sizes larger than observed ones, before fragmenting. We show that including magnetic field in these simulations reduces the size of the outcoming disks, and prevents them from fragmenting, as showed by previous studies.}
    {%Conclusions:
    %Protostellar disks can emerge from an asymmetrical gravitational collapse of a dense core. The order of magnitude of the specific angular momentum is entirely compatible with observed values. \textcolor{SeaGreen}{(To be confirm by Anaelle, not in abstract?) The orientation of apparent velocity gradients in these simulations seems to match with recent observations.}
    We show that in a non-axisymmetrical collapse, the formation of a disk can be induced by small perturbations of the initial density field in the core, even in the absence of global large-scale rotation of the core. In this scenario, large disks are generic features which are natural consequences of the hydrodynamical fluid interactions and self-gravity. Since recent observations have shown that most disks are significantly smaller having a size of a few tens of astronomical units, our study suggests that magnetic braking is the most likely explanation. The kinematics of our model are consistent with typically observed values of velocity gradients and specific angular momentum in protostellar cores. These results open a new avenue to explore in our understanding of the early phases of disk formation, since they suggest a fraction of the protostellar disks could be the product of non-axisymmetrical collapse and not resulting from the conservation of large scale angular momentum in rotating cores.}

  %\keywords{Methods: analytical - Methods: numerical - ISM: clouds - ISM: magnetic fields - Turbulence - Stars: formation}

  \maketitle

%============================================================================
\section{Introduction}
Protoplanetary disks are rotationally supported structures that formed around young stars \citep{Li2014, Dutrey2014, Testi2014}. It is currently believed that the rotation of these disks is inherited from large scales -- a few thousands of astronomical units, the scale of the parent prestellar dense core.

During the gravitational collapse of the core, if the angular momentum is conserved, the infalling material naturally forms a rotation dominated structure at the small scale of a hundred astronomical units. The rotation of the disk is thus inherited from the large scale angular momentum, and as a consequence, the velocity gradients at large and small scales are correlated. This scenario is extensively studied in the literature and in particular a majority of collapse calculations starts with a prescribed rotation profile (see for example \citet{Bate1998,Matsumoto2003,Machida2005,Hennebelle2008}). While reasonable, this scenario leads to the question at which scale is the angular momentum being inherited from and how exactly this happens. Another frequently configuration consists in a cloud with a turbulent velocity field imprinted initially \citep{Bate2003,Goodwin2004A,Goodwin2004,Dib2010,Hennebelle2016,Gray2018,Kuznetsova2019}. In this context as well, it has been found that disks quickly form. The usual interpretation is that the angular momentum is initially present because of the turbulence.

Observationally, the kinematics of the dense gas in both prestellar cores and protostellar envelopes has been studied thanks to the analysis of molecular line emission and shown to harbor velocity gradients at scales \SI{0.01}{} to \SI{0.1}{pc}, interpeted as rotation of cores (see the early works by \citet{Goodman93, Ohashi1997, Chen2007}). The values of specific angular momentum measured inside protostellar envelopes at scales a few thousand astronomical units are on average one order of magnitude lower than the ones observed at larger scales in starless structures (a few \SI{e-4}{km.s^{-1}.pc}, see for example \citet{Belloche2013}, \citet{Yen2015a} and the very recent work by \citet{Gaudel2020}). More puzzling, however, were the observations showing that some protostellar cores show an apparent disorganization or even reversal of their velocity pattern, sometimes interpreted as the contribution of infall motions to the projected velocity field \citep{Tobin2011, Harsono2014} or as counter rotation \citep{Tobin2018,Takakuwa2018}. Recent high dynamic range observations of a sample of 12 low-mass Class 0 protostars (in the {\sc calypso} sample) by \citet{Gaudel2020} exhibit systematic dispersion of velocity gradients between disk's and envelope's scales, questioning the presence of large scale rotation. Moreover, observations of the specific angular momentum contained in T-Tauri disks suggest values larger, by about one order of magnitude, than the specific angular momentum observed in the low-mass protostellar cores at scales a few thousands astronomical units \citep{Simon2000,Kurtovic2018,Perez2018}. These observations are hence difficult to reconcile with a simple picture of rotating-infalling protostellar envelope which conservation of angular momentum naturally produces a rotationally-supported disk in its center, and new models should be developed to reproduce these observations as well.

To tackle these issues we investigate a scenario that also leads to the formation of a protostellar disk. Similar ideas than the ones exposed in this paper has already been developed in the context of spiral galaxies formation by \citet{Hoyle49, Sciama55, Peebles69}. In the context of protoplanetary disk formation this paper is meant to be exploratory, so we used minimal physical ingredients. Our simulations are thus purely hydrodynamics, except in part \ref{part_MHD}.
We start from an extreme scenario, considering a perfectly motionless dense core with non-axisymmetric density perturbations.
%This scenario starts with a motionless dense core with non-axisymmetric density perturbations. 
The gravitational collapse of this core is thus non-axisymmetric. We show analytically and numerically that this non-axisymmetry leads to the possibility to generate rotation locally. 
As we start from an extreme motionless scenario, without considering velocity fluctuations, our model is not fully physical. Despite this fact, we then analyse the velocity gradients in our simulations and we reproduce the observational results from \citet{Gaudel2020} about the dispersion of velocity gradients. We show that the specific angular momentum step coincides with \citet{Belloche2013} and \citet{Gaudel2020} results. Our model thus exhibit good agreement with observational constraints on kinematics.

The plan of the paper is as follows: in the second part we present the theory that motivates our study, in the third part we present the numerical methods we used to investigate our problem, in the fourth and fifth parts we present the results obtained and their discussion, and the sixth part is the conclusion.

\section{Theory}
\label{theory_part}
\subsection{The axisymmetrical case}
In the introduction, we evoked the \emph{conservation of angular momentum} during the collapse of a dense core, leading to the formation of a disk. However, the angular momentum is correctly defined only in a given frame and with respect to a given point. Let's consider the angular momentum calculated in the simulation box frame $\mathcal{R}$, with respect to the center $O$ of this box\footnote{In fact, in relation to any fixed point of the simulation.}. It is computed as follows, with the summation referring to the different cells $i$ of the simulation, $m_i$ and $M_i$ being respectively the mass and position of each cell $i$.

\begin{equation}
\left. \bm{\sigma_O} \right| _{_{\mathcal{R}}} = \sum_i m_i \bm{OM_i} \times \frac{\d \bm{OM_i}}{\d t}
\label{moment 0}
\end{equation}

\noindent
This momentum is conserved in virtue of the fundamental law of evolution of the angular momentum in a Galilean frame:

\begin{equation}
\frac{\d \left. \bm{\sigma_O} \right| _{_{\mathcal{R}}}}{\d t} = \sum_i \bm{\mathcal{M}_O} \left( \bm{F}_{\text{ext} \rightarrow i} \right) = \bm{0}
\label{eq evol moment 0}
\end{equation}

\noindent
As no external force is applied on the system, the angular momentum $\left. \bm{\sigma_O} \right| _{_{\mathcal{R}}}$ is conserved. In a simple axisymmetrical case, this momentum coincides with what we will call \emph{the momentum of the disk}. Indeed, during the collapse, a disk forms in the center of the box, thus $\left. \bm{\sigma_O} \right| _{_{\mathcal{R}}}$ represents the angular momentum computed in the frame of the disk, in relation to the center of the disk.

\subsection{Non-axisymmetric configuration}
\label{formation disque theorique}
In a non-axisymmetrical case, $\left. \bm{\sigma_O} \right| _{_{\mathcal{R}}}$ is no longer a relevant quantity to study the disk formed in the simulation. In fact, to measure the angular momentum in protostellar disks, the reference point with respect to which the angular momentum is computed is the center of the disk, and the velocities considered are those in relation to the center of the disk, deducted from those projected on the line of sight \citep{Belloche2013}. In the axisymmetrical case, the center of the formed disk remains motionless at the center of the simulation box. In the non-axisymmetrical case, the disk is not formed at the center of the simulation box, and have a proper motion. We thus have to consider the angular momentum computed in the frame $\mathcal{R'}$ of the disk, in relation to the center $C$ of the disk:

\begin{equation}
\left. \bm{\sigma_C} \right| _{_{\mathcal{R'}}} = \sum_i m_i \bm{CM_i} \times \frac{\d \bm{CM_i}}{\d t}
\label{moment reel}
\end{equation}

We show in appendix \ref{dev_theor_part_1} that for an initial condition where all cells are at rest in $\mathcal{R}$, $\left. \bm{\sigma_C} \right| _{_{\mathcal{R'}}}$ can be expressed as:

\begin{equation}
\left. \bm{\sigma_C} \right| _{_{\mathcal{R'}}} =  M \bm{GC} \times \frac{\d \bm{GC}}{\d t} =  M \bm{GC} \times \frac{\d \bm{OC}}{\d t}
\label{eq moment dans R'}
\end{equation}

\noindent
where $M = \sum\limits_i m_i$ is the total mass of the system, and $G$ is the center of mass\footnote{The center of mass remains motionless in the frame $\mathcal{R}$ of the simulation box due to the lack of external force. See appendix \ref{dev_theor_part_1} for more details.}.% such as:

%\begin{equation}
%\bm{CG} = \frac{1}{M} \sum_i m_i \bm{CM_i}
%\end{equation}

\noindent
Furthermore, the time derivative of Eq. (\ref{eq moment dans R'}) gives:

\begin{equation}
\frac{\d \left. \bm{\sigma_C} \right| _{_{\mathcal{R'}}}}{\d t} = M \bm{GC} \times \frac{\d ^2 \bm{GC}}{\d t^2}
\label{eq cons moment dans R'}
\end{equation}

\noindent
This equation can be interpreted as the variation of the angular momentum $\left. \bm{\sigma_C} \right| _{_{\mathcal{R'}}}$ due to the torques of inertial force that apply on each cell of the simulation in the non Galilean frame $\mathcal{R'}$ (see appendix \ref{inertial_force_development} for detailed development). As the center of mass $G$ and the center of the formed disk $C$ do not coincide, the angular momentum $\left. \bm{\sigma_C} \right| _{_{\mathcal{R'}}}$ is not expected to be conserved.

The point $C$ being the accretion center of the system, the matter collapses toward it. As the angular momentum (in $\mathcal{R'}$ with respect to $C$) of this matter does not vanish, a rotationally supported structure forms around $C$.

In the axisymmetrical case, the three points $C$, $G$ and $O$ coincide, as well as the two frames $\mathcal{R}$ and $\mathcal{R'}$. Thus $\left. \bm{\sigma_C} \right| _{_{\mathcal{R'}}} = \left. \bm{\sigma_O} \right| _{_{\mathcal{R}}}$ and the angular momentum is therefore conserved through the temporal evolution of the structure. In the non-axisymmetrical case, Eq. (\ref{eq moment dans R'}) shows that the angular momentum $\left. \bm{\sigma_C} \right| _{_{\mathcal{R'}}}$ is equal to the angular momentum of the point $C$ to which the whole simulation mass has been allocated, computed with respect to $G$, in the frame $\mathcal{R}$ of the simulation box. Eq. (\ref{eq cons moment dans R'}) shows that this angular momentum is not conserved in the general case. We consider here the extreme case where every cells are initially at rest in the simulation frame $\mathcal{R}$. $\left. \bm{\sigma_C} \right| _{_{\mathcal{R'}}}$ is thus initially null, $C$ being the accretion center of the system. As matter is falling toward the center, the transversal velocity of this matter has to increase for the angular momentum to grow. As a result, a rotational structure naturally forms around the accretion center, without violating the conservation of the angular momentum in the simulation frame $\mathcal{R}$. These ideas are similar than the ones of \citet{Peebles69} who showed that the angular momentum of spiral galaxies can be understood as a consequence of tidal torques acting during the gravitational collapse.

In our context of protostellar disk formation and before analysing in details our simulations in part \ref{analyse 50vhr}, we can provide an order of magnitude for the expected momentum $\left. \sigma_C \right| _{_{\mathcal{R'}}}$. The simulations show that the distance between the points $C$ and $G$ is about a thousand astronomical units, which corresponds roughly to a tenth of the dense core radius --- \SI{875}{AU}. Let's take for $\left| \left| \frac{\d \bm{GC}}{\d t} \right| \right|$ the order of magnitude of the sound speed, \SI{0.2}{km.s^{-1}}. The total mass inside the simulation being \SI{2.5}{\Msun}, we obtain:

%\[
%\left. \sigma_C \right| _{_{\mathcal{R'}}} \sim 2.5 \Msun ~.~ 10^3 AU ~.~ 0.2 km.s^{-1} \sim 3.10^{47} kg.m^2.s^{-1}  
%\]

\[
\left. \sigma_C \right| _{_{\mathcal{R'}}} \sim \SI{2.5}{\Msun} ~.~ \SI{e3}{AU} ~.~ \SI{0.2}{km.s^{-1}} \sim \SI{e47}{kg.m^2.s^{-1}}  
\]

\noindent
As will be seen later, this order of magnitude is consistent with the values given by our analyses (see Fig. \ref{analyses erreurs et disque}) and stresses good agreement with \citet{Belloche2013} and \citet{Gaudel2020}.

\noindent
\begin{figure}
%\resizebox{\linewidth}{!}{\includegraphics{Figures/B335_noturb_norot_hydro_pert_asym_aleatoire_shr_bigbox_50pourc_dens_x_6_1.pdf}}
\includegraphics{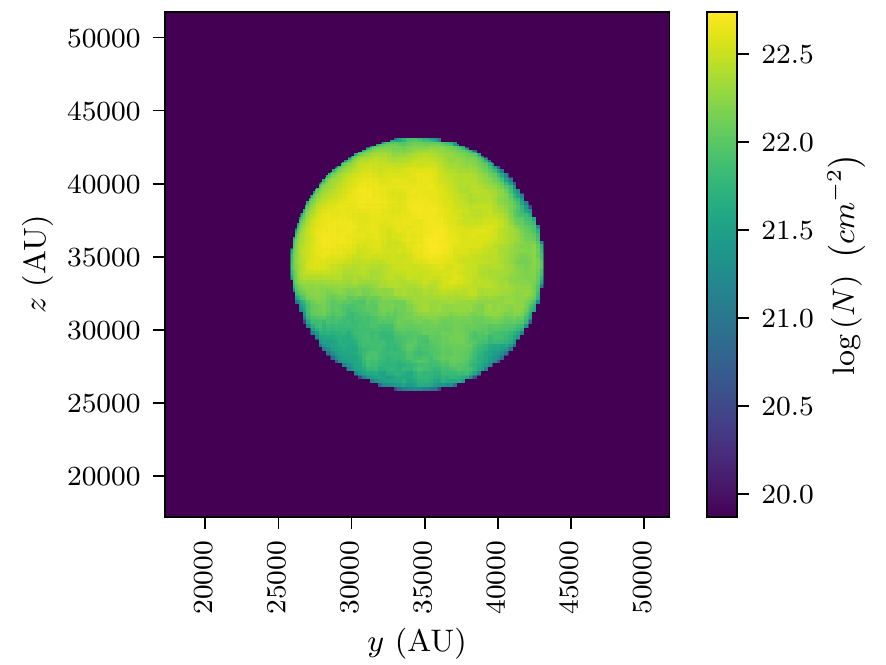}
\caption{Column density map along the $y$ axis of the simulation box, representing the initial conditions with a perturbation level of $50 \%$. These perturbations in density are based on a Kolmogorov turbulent spectrum. Initially all the velocities are set to zero so that the angular momentum in the box frame is initially null.}
\label{condinit pert}
\end{figure}

\noindent
\begin{figure*}
\begin{center}
\resizebox{0.4\linewidth}{!}{\includegraphics[width=\textwidth]{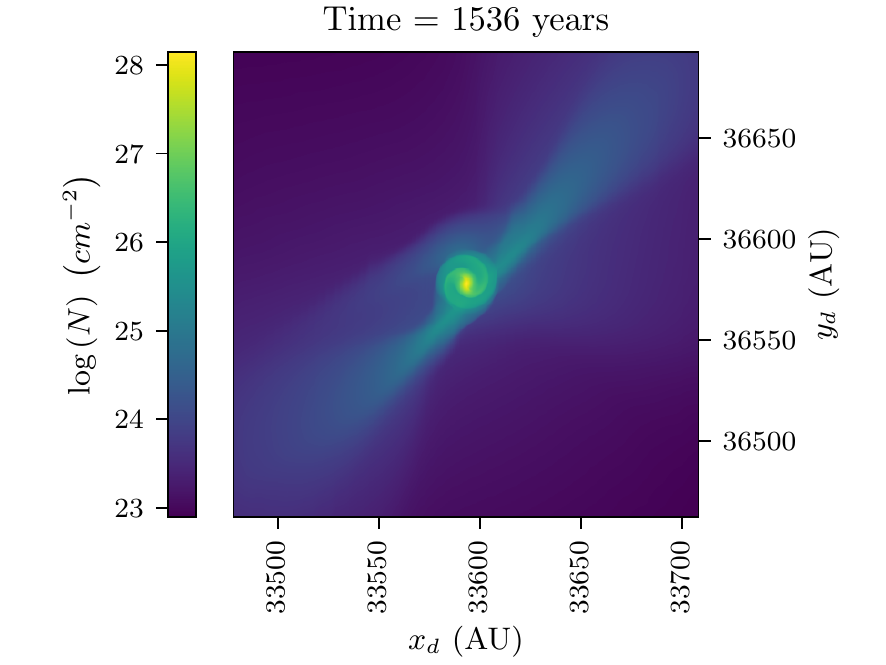}}%
\resizebox{0.4\linewidth}{!}{\includegraphics[width=\textwidth]{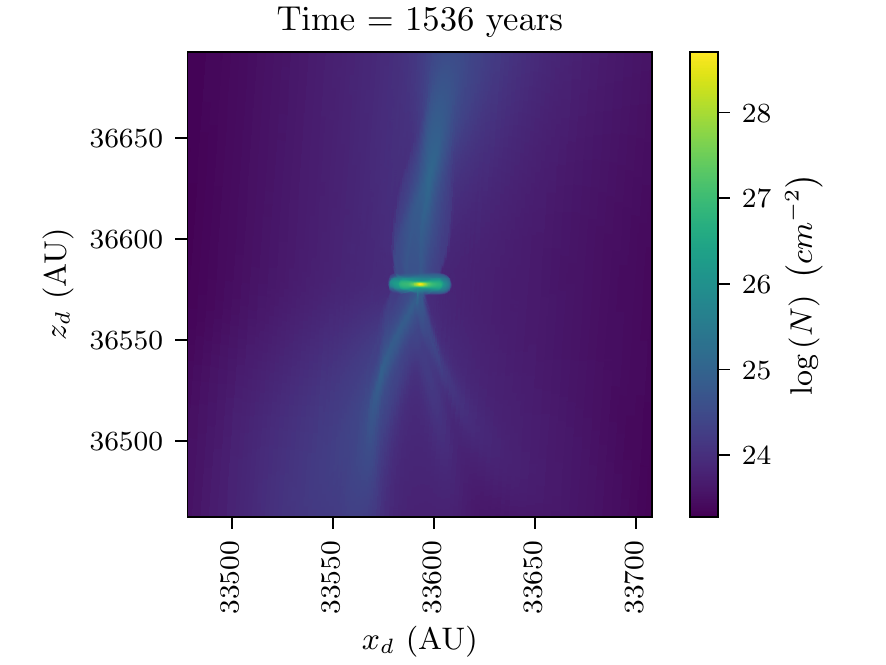}}
\resizebox{0.4\linewidth}{!}{\includegraphics[width=\textwidth]{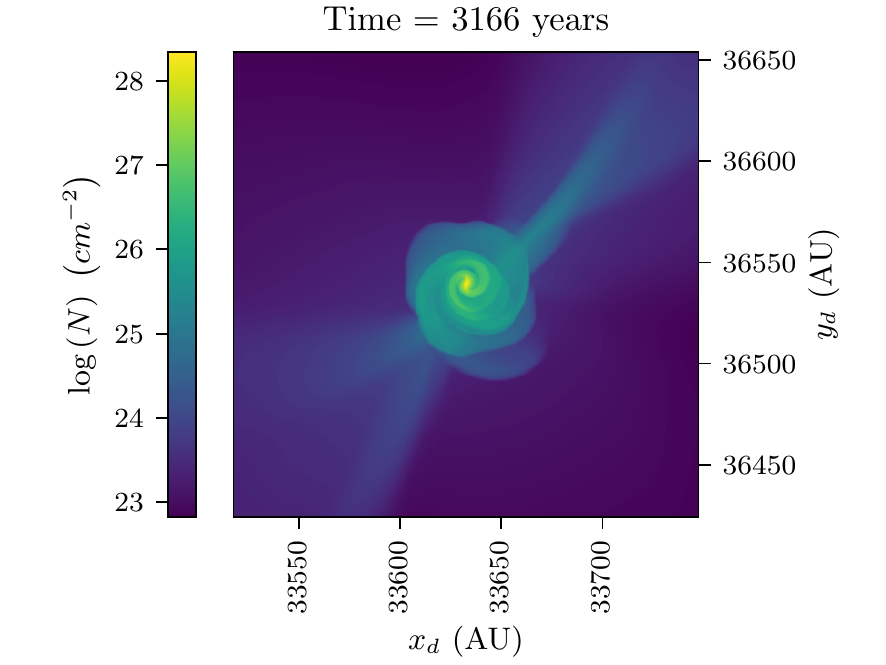}}%
\resizebox{0.4\linewidth}{!}{\includegraphics[width=\textwidth]{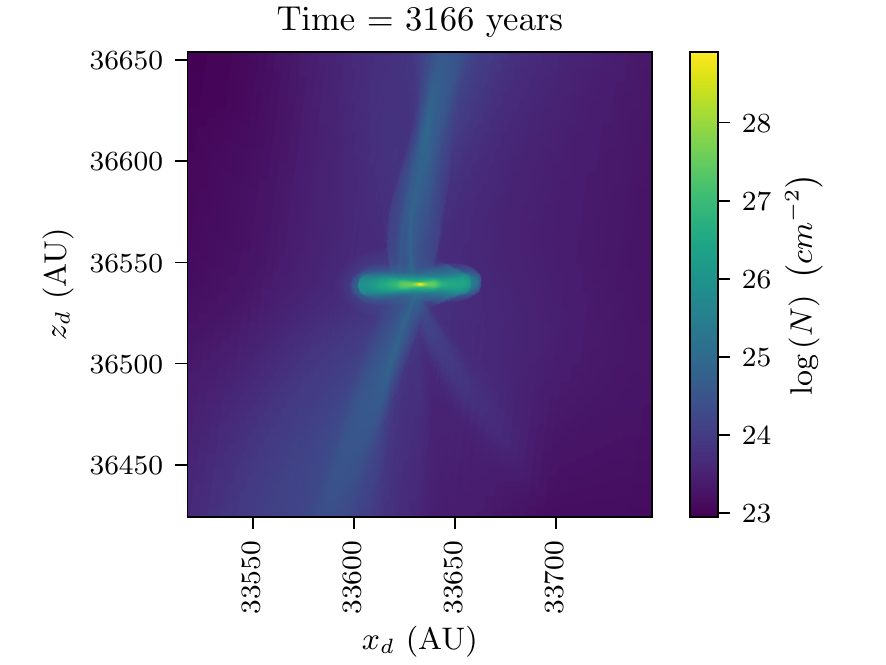}}
\resizebox{0.4\linewidth}{!}{\includegraphics[width=\textwidth]{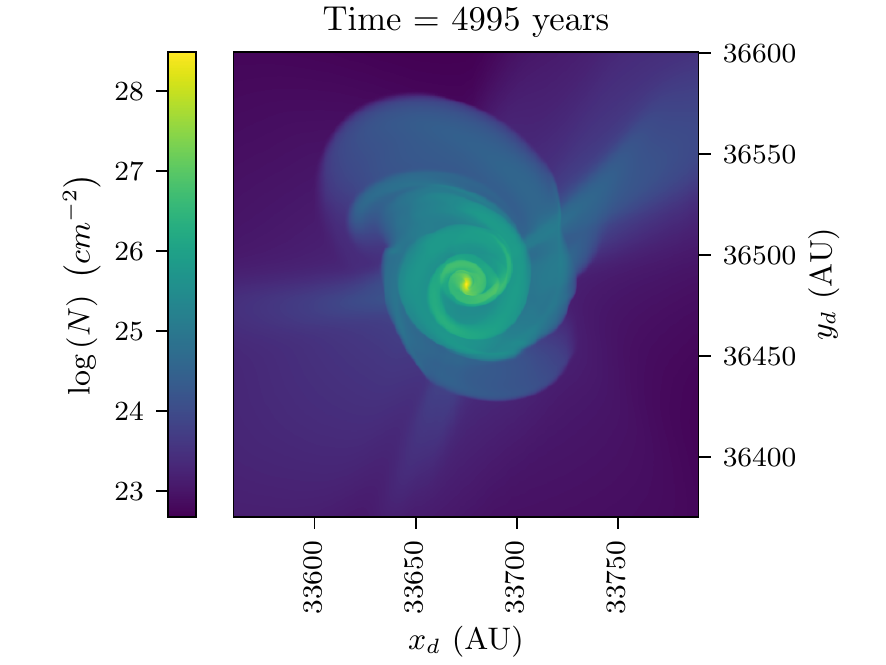}}%
\resizebox{0.4\linewidth}{!}{\includegraphics[width=\textwidth]{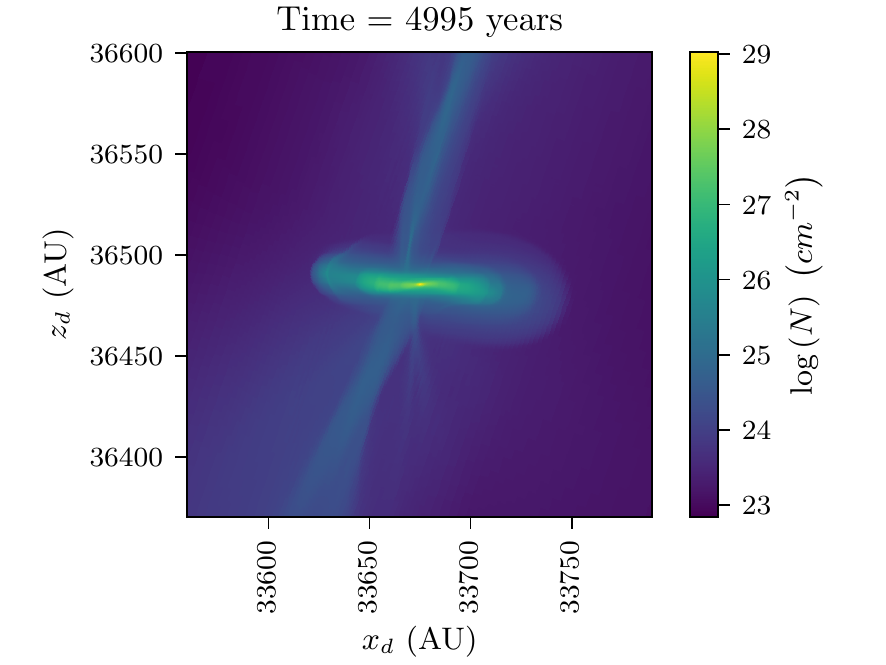}}
%%\caption{Column density (on the left) and projected %velocity along the line of sight (on the right), along %the $x$ axis of the simulation box, for the simulation %with $50 \%$ of perturbations. We see that a disk forms %and grows in spite of the fact that initially the %angular momentum is null.}
\caption{Column density in the simulation with $50 \%$ of perturbations at three different times. \emph{Left:} face-on projection. \emph{Right:} edge-on projection. The rotation axis $z_d$ of the disk is not overlapping with any simulation box axis. We see that a disk forms and grows in spite of the fact that initially the angular momentum is null.}
\label{sequence disk}
\end{center}
\end{figure*}

\section{Numerical methods}
\label{partie_erreur}

\subsection{Code and numerical parameters}
To study whether the asymmetry of a gravitational collapse can be sufficient to form a prestellar disk, we carry out a set of hydrodynamics simulations with \ramses{} \citep{Teyssier2002}. This numerical Eulerian code uses Adaptative Mesh Refinement (AMR) technique to enhance resolution locally, where it is needed, on a Cartesian mesh. Our refinement criterion is based on Jeans length such as each local Jeans length is described by at least 40 cells. We use ten levels of AMR, from 8 to 18, leading to a spatial resolution which goes from \SI{270}{AU} to \SI{0.26}{AU}.

\subsection{Initial conditions}
\label{def pert}
We consider a 3D cubic box with sides of \SI{0.33}{pc} (about \SI{70000}{AU}). At the middle of the box we place a \SI{2.5}{\Msun} sphere of gas which diameter is the quarter of the box length and which radial density profile is flat. This sphere acts as a model of prestellar dense core. The rest of the box is filled with an envelop of gas whose density is constant and equal to a thousandth of the mean density of the gas sphere. Initially all velocities are set to zero so that the angular momentum in the box frame with respect to any motionless point is initially null. The alpha parameter -- thermal over gravitational energy ratio -- of the gas sphere is \SI{0.35}{}. We use a barotropic equation of state for the gas:

\begin{equation}
T=T_0 \left( 1+ \left( \frac{\rho}{\rho_c} \right) ^{\gamma -1} \right)
\end{equation}

\noindent
$T$ and $\rho$ being the temperature and density of the gas, $T_0 = \SI{10}{\kelvin}$, the critical density $\rho_c = \SI{e-13}{g.cm^{-3}}$ and the adiabatic index $\gamma = \SI{1.4}{}$.

As we want to study the effect of asymmetrical gravitational collapse, we break the symmetry of the cloud introducing random density perturbations in the dense core. To mimic roughly the physics of the interstellar gas, we based the probability distribution of the density perturbations on the one of the turbulence (see for example chapter 3 of \citet{Hennebelle2012} for a review on turbulence in interstellar clouds). In the Fourier space the perturbations spectrum matches with the power spectrum of the velocity for Kolmogorov turbulence\footnote{$\mathcal{P}_v (k) \propto k^{-11/3}$}, and the phases are randomly chosen. We can thus write the density of each cell $i$ of the prestellar core as :

\begin{equation}
d_i = d_0 (1 + A . \delta \rho_i)
\label{eq_dens}
\end{equation}

\noindent
where $d_0$ represents the mean density of the prestellar core, $\delta \rho_i$ is the value of the perturbation at the considered point, and $A \in \left[ 0, ~1 \right] $ is an internal parameter allowing us to control the amplitude of the perturbation. To ensure that the density stay positive everywhere, it is necessary that $~~\delta \rho_i \in \left] -1, ~+ \infty \right[~$. To satisfy this condition, we modified all the values of $\delta \rho_i < -1$ to bring them to \SI{-0.99}{}.

%\noindent
%\begin{figure}
%\begin{center}
%\resizebox{\linewidth}{!}{\includegraphics{Histogramme_perturbation_RMS}}
%\end{center}
%\caption{Histogram of the density perturbations of the prestellar core.}
%\label{hist pert}
%\end{figure}

To assure that the mean density of the prestellar core stay constant when vaying $A$, we renormalised\footnote{The operation is simply : $d_{i,\text{\emph{af}}} = d_{i,\text{\emph{be}}} \frac{d_0}{<d_{i,\text{\emph{be}}}>_i}$ with $\text{\emph{be}}$ and $\text{\emph{af}}$ refering to \emph{before} and \emph{after} the renormalisation operation, which naturally leads to $<d_{i,\text{\emph{af}}}>_i = d_0$.} the value of $d_i$ defined in Eq. (\ref{eq_dens}) in each cell : 

\begin{equation}
<d_i>_i  = d_0
\end{equation}

\noindent
where the operator $<.>_i$ represents the mean value over all the cells $i$ of the prestellar dense core. As the mean value of $\delta \rho_i$ is not null, this operation warrants that we can change the amplitude of the perturbations without modifying the mean value of the density.

Least, we call \emph{perturbation level} the ratio between the root mean square value of the perturbations and the mean value of density, that we express as a percentage :

\begin{equation}
\varepsilon = 100 \frac{\sqrt{\left< \left( d_i - d_0 \right) ^2 \right>_i}}{d_0}
\end{equation}

An example of initial conditions including density perturbations constructed as described above is presented in Fig. \ref{condinit pert}. We stress that unlike most of previous dense core collapse studies, which attempted to form and study disks, we have no rotation or turbulence initially.

%\noindent
%\begin{figure}
%\resizebox{\linewidth}{!}{\includegraphics{Figures/B335_%noturb_norot_hydro_pert_asym_aleatoire_shr_bigbox_50pourc%_dens_x_6_1.pdf}}
%\includegraphics{Figures/B335_noturb_norot_hydro_pert_asy%m_aleatoire_shr_bigbox_50pourc_dens_x_6_1.pdf}
%\caption{Column density map along the $y$ axis of the %simulation box, representing the initial conditions with %a perturbation level of $50 \%$. These perturbations in %density are based on a Kolmogorov turbulent spectrum. %Initially all the velocities are set to zero so that the %angular momentum in the box frame is initially null.}
%\label{condinit pert}
%\end{figure}

\subsection{Choice of a time reference}
Since the free-fall time depends on the density, it can slightly vary from a level of perturbation to another. As the grid is initially coarse, the first time-steps of the simulation are much larger than the later ones after collapse occurred, when the level of refinement is higher. These two effects lead to a bad description in time at the beginning of the simulation. We are thus compelled to choose a time reference from which the ages are computed. We based our time reference on the maximum density inside the simulation box. We take for time reference the moment when the maximum density reaches \SI{e-13}{g.cm^{-3}}. It corresponds to the limit density beyond which the compression of the dense gas changes from isothermal to adiabatic \citep{Larson69}.

\section{Results}
\label{partie_pert}

\subsection{Formation of a disk}
The main result of this study is, as we expected from our theoretical development, the formation of a disk within our simulations. We show a sequence of images of the disk growing with time in the simulation with $50 \%$ of density perturbations in Fig. \ref{sequence disk}. This disk is very similar %reminiscent
to the ones found in many studies \citep{Bate2003,Matsumoto2003, Goodwin2004A, Hennebelle2008, Gray2018}. As can be seen it presents prominent spiral arms that transport angular momentum. To verify that this structure is a \emph{disk}, we made sure to check that the formed structure is rotationally supported. In order to keep our model as simple as possible, we did not use sink particles \citep{Krumholz2004,Bleuler2014} in our simulations presented in this part and analysed in parts \ref{analyse 50vhr} and \ref{part_velocity_gradients}. As a consequence, the dense gas accumulates in the disk, making it self-gravitating. Thus the azimuthal velocity profile shows substantial deviations from the keplerian profile, but the structure is still rotationally supported, as the azimuthal velocity is much larger than the radial velocity inside the disk, typically by a factor $50$. %In this condition, to demonstrate that this structure is rotationally supported, we computed the orthoradial velocity as a function of the radius from the center of the structure and showed that it is much larger than the radial velocity inside the disk, typically by a factor $50$.
We have however also showed that disks form in simulations with sink particles (see part \ref{simu_sink}) and we verified that the formed structures match very well the keplerian profile.

We run a set of simulations which differ only by their perturbation level, from $10\%$ to $60\%$. In all of these simulations we observe the formation of a protostellar disk. The comparison with theory, studied in details for the simulation with $50\%$ of density perturbations in part \ref{analyse 50vhr} gives clues to assess the trust level of our simulations (for the other levels of perturbation, the results are visible in appendix \ref{annexe conv num}). For the simulations over $20\%$ of perturbations, the comparison showed that the formation of the disk can be trusted. For the simulations with low perturbation level, typically less than $20\%$, the agreement with theory is less good and the numerical errors are higher.

%\noindent
%\begin{figure}
%\resizebox{0.5\linewidth}{!}{\includegraphics[width=\textwidth]{Figures/B335_noturb_norot_hydro_pert_asym_aleatoire_shr_bigbox_50pourc_dens_x_3_59.pdf}}%
%\resizebox{0.5\linewidth}{!}{\includegraphics[width=\textwidth]{Figures/B335_noturb_norot_hydro_pert_asym_aleatoire_shr_bigbox_50pourc_vel_x_3_59.pdf}}

%\caption{Column density (on the left) and projected velocity along the line of sight (on the right), along the $z$ axis of the simulation box, for the simulation with $50 \%$ of perturbations \textcolor{DarkMagenta}{(changer les images pour la vraie simulation à 50\%, en bigbox, ne pas mettre le temps corrigé et mettre le temps dans le cadre, pas au-dessus, agrandir les légendes des axes)}.}
%\label{disque 50hr}
%\end{figure}

\subsection{Comparison with theory}
\label{analyse 50vhr}
We showed in part \ref{formation disque theorique} that a rotational structure should emerge from an asymmetric gravitational collapse, even without any initial motion. To compare this theoretical prediction with our simulations, we exploit Eqs. (\ref{moment reel}) and (\ref{eq moment dans R'}). The first expression simply expresses the numerical way to compute the total angular momentum in the simulation (in the frame $\mathcal{R'}$, with respect to the point $C$, what we will not mention anymore). Let's name this quantity $\bm{\sigma_\text{num}}$. The second one is the analytic expression of the same quantity. It is equivalent to the first one if a set of hypothesis is verified (see appendix \ref{dev_theor_part_1}), which is the case in our simulations. Let's name this quantity $\bm{\sigma_\text{an}}$. The equality of the two quantities $\bm{\sigma_\text{num}}$ and $\bm{\sigma_\text{an}}$ ensures that we are observing the physical phenomenon we described. The differences could be due to numerical errors. In the first place, we verify numerically the equality:

\begin{equation}
\begin{split}
%\bm{\sigma_\text{num}} ~~ & =~~~ \sum_i m_i \bm{CM_i} \wedge \frac{\d \bm{CM_i}}{\d t} \\  & \overset{?}{=}~~~ M \bm{GC} \wedge \frac{\d \bm{OC}}{\d t} ~~=~~ \bm{\sigma_\text{an}}
\bm{\sigma_\text{num}} ~ & =~ \sum_i m_i \bm{CM_i} \wedge \frac{\d \bm{CM_i}}{\d t}  ~~~~\overset{?}{=}~~~~ M \bm{GC} \wedge \frac{\d \bm{OC}}{\d t} ~=~ \bm{\sigma_\text{an}}
\end{split}
\end{equation}

\begin{figure}
\begin{center}
\resizebox{\linewidth}{!}{\includegraphics{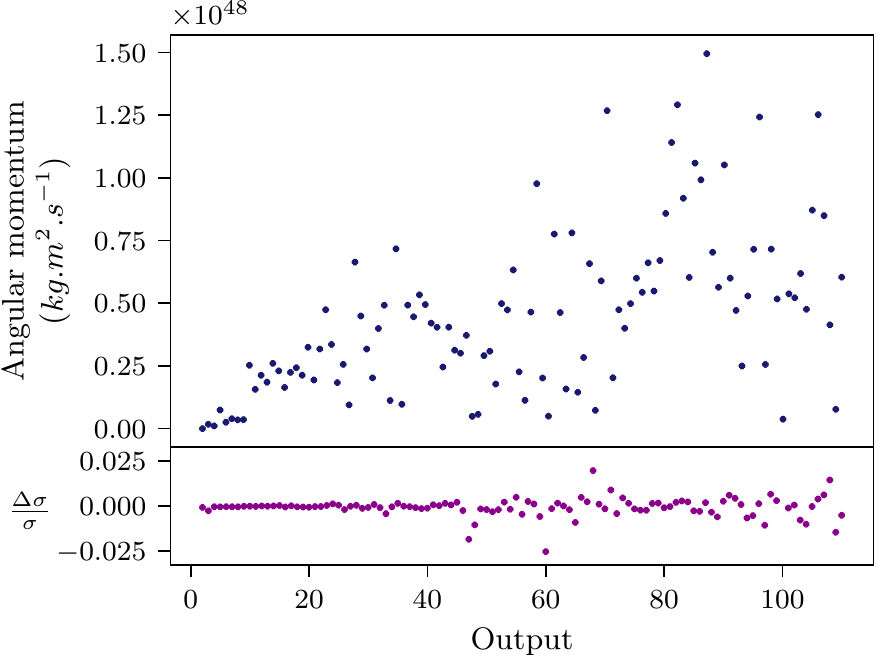}}

\resizebox{\linewidth}{!}{\includegraphics{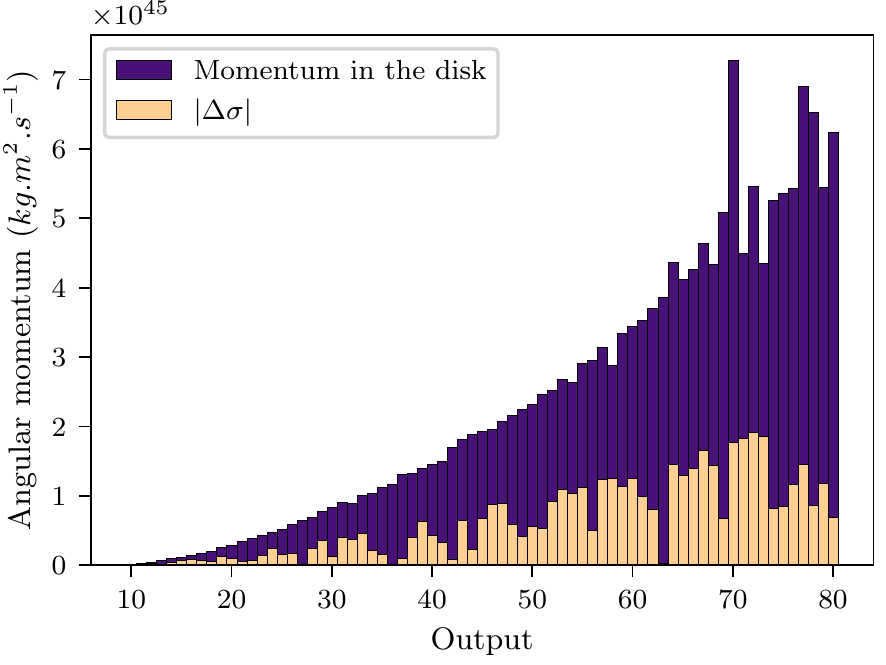}}
\end{center}
\caption{Analysis of the simulation with $50\%$ of perturbations. \emph{Top}: $\sigma_\text{num}$ (see Eq. (\ref{moment reel})) in blue, and the relative difference between $\sigma_\text{num}$ and $\sigma_\text{an}$ (see Eq. (\ref{eq moment dans R'})) in purple for each output. \emph{Bottom}: the momentum in the disk (in violet) compared to the absolute value of the difference between $\sigma_\text{num}$ and $\sigma_\text{an}$ (in beige). We see that $\frac{\Delta \sigma}{\sigma}$ is less than $2.5\%$ and that this difference is smaller than the momentum contained in the disk. The disk is thus resulting in the physics we described part \ref{formation disque theorique}.}
\label{analyses erreurs et disque}
\end{figure}

The comparison between $\sigma_\text{num}$ and $\sigma_\text{an}$ is represented on the top panel of Fig. \ref{analyses erreurs et disque} for the simulation with $50 \%$ of perturbations. The slight relative difference --- globally less than $2\%$ ---  %seems to confirm the equality of these two quantities. 
shows that the equality of these two quantities is numerically consistent.
Furthermore, the relative difference does not keep the same sign during the temporal evolution. It shows that this difference is not a systematic error on one of the two quantities.
%As this relative difference switches sign during the temporal evolution, it ensures that we are not facing an accumulation of numerical errors.

\noindent

Let's name $\left| \Delta \sigma \right| = \left| \sigma_\text{num} - \sigma_\text{an} \right|$. To be entirely sure that the disk formed in our simulation is not a numerical artifact, we verified that the angular momentum it contains is larger than $\left| \Delta \sigma \right|$. In the most pessimistic scenario where $\left| \Delta \sigma \right|$ would be entirely concentrated in the disk, this ensures that $\left| \Delta \sigma \right|$ is not sufficient to explain the presence of the disk.
The angular momentum contained in the disk is computed as $\sigma_\text{num}$, but in the restricted area of the simulation corresponding to the disk\footnote{To belong to the disk, we consider that a cell has to have a high enough density and to have an azimuthal velocity larger than twice the radial velocity.}.
The comparison between $\left| \Delta \sigma \right|$ and the angular momentum in the disk is visible on the bottom panel of Fig. \ref{analyses erreurs et disque}. As $\left| \Delta \sigma \right|$ is smaller than the angular momentum in the disk, and as $\Delta \sigma$ is switching sign over the temporal evolution of the system, it confirms that the formed disk is the result of the physics described part \ref{formation disque theorique}. For the other levels of perturbations, the results are presented in appendix \ref{annexe conv num}.

\subsection{Analysis of velocity gradients}
\label{part_velocity_gradients}
In this section we analyse our simulations from an observational point of view to highlight whether or not our model succeed to describe some features of real observations. As the rotation emerges from the asymmetrical gravitational collapse in our model, it is not straightforwardly inherited from larger scales. This lack of connection between disk and envelope scales leads to important features. We compute three quantities at different scales that are accessible to observations: the direction and amplitude of velocity gradients and the specific angular momentum.

To compute direction and amplitude of the velocity gradients at a scale $R$ we follow the method of \citet{Goodman93}. We consider a cube of side $2R$ around the disk, aligned with the three main axis of the simulation. We consider these three main axis as our lines of sight to compute maps of projected velocities weighted by density, on a depth of $2R$. Then we fit these maps with a solid-body rotation profile:

\begin{equation}
    v_{LSR} = v_0 + a \Delta x + b \Delta y
    \label{solid-body rotation}
\end{equation}

\noindent
with $v_0$ the systemic velocity of our object, $\Delta x$ and $\Delta y$ the vertical and horizontal dimensions of our projected velocities map. The magnitude of the velocity gradient is thus $\Omega = \sqrt{a^2 + b^2}$ and its direction is given by $\theta = \arctan \frac{a}{b}$. The specific angular momentum at a scale $R$ is thus given by $j=R^2 \Omega$.

\begin{figure}
\begin{center}
\resizebox{\linewidth}{!}{\includegraphics{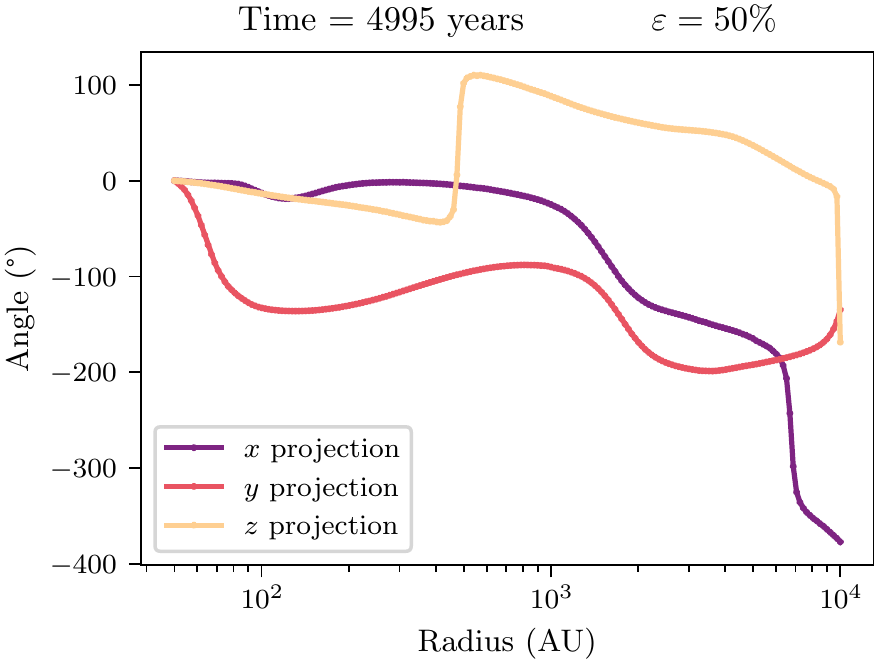}}

\resizebox{\linewidth}{!}{\includegraphics[trim = 0cm 0cm 0cm 0.3cm, clip]{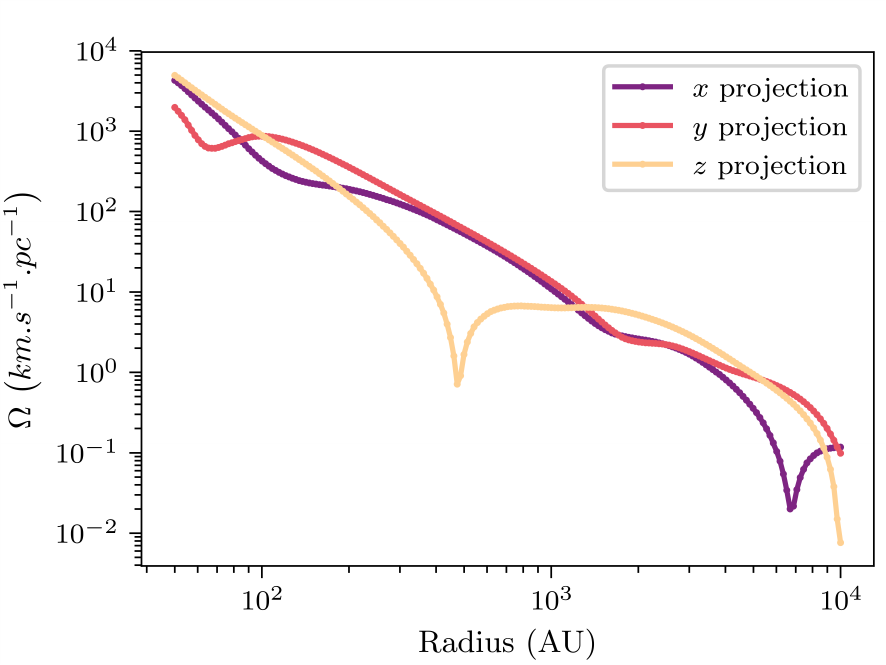}}

\resizebox{\linewidth}{!}{\includegraphics[trim = 0cm 0cm 0cm 0.3cm, clip]{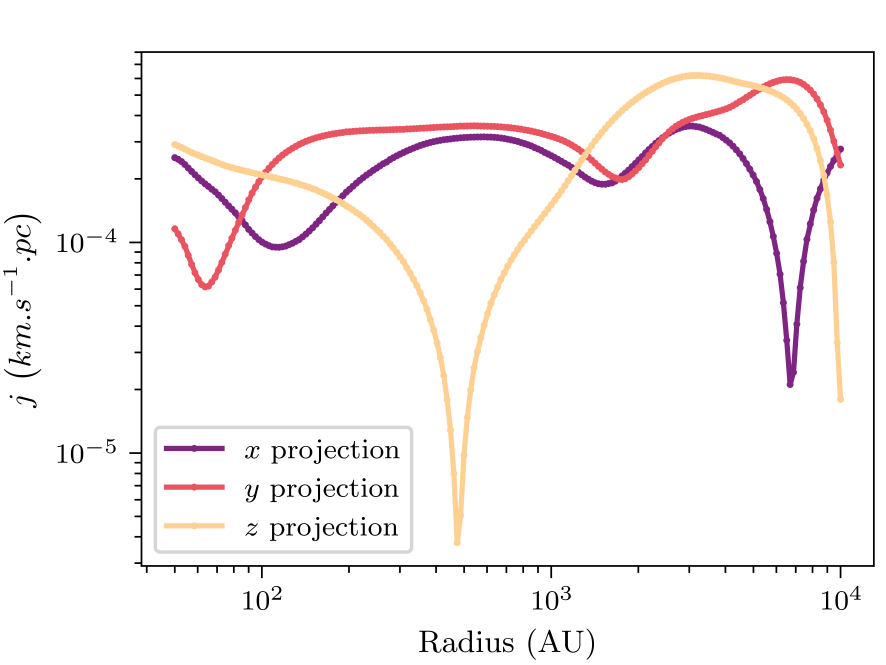}}
\end{center}
\caption{Analysis of velocity gradients at different scales in the simulation with $50\%$ of perturbations. On each panel, the three curves correspond to the three main projections of the simulation. \emph{Top}: angular direction of velocity gradients. The origin of the angular direction corresponds to the direction of the disk scale gradient. \emph{Mid}: amplitude of velocity gradients. \emph{Bottom}: specific angular momentum as computed in observation analysis.}
\label{gradients 50shr}
\end{figure}

In an axisymmetrical model with initial rotation, velocity gradients at small and large scales are perfectly aligned. In our model, due to the lack of initial rotation, it is interesting to see how velocity gradients at different scales are organised. The top panel of Fig. \ref{gradients 50shr} shows the angular variation of velocity gradients according to the probed scale, relatively to the velocity gradient at the disk scale. Velocity gradients in the disk and in the envelope are misaligned. For the $y$ projection, the velocity gradient in the envelope is even reversed in comparison to the small scale gradient, whereas for the $x$ projection the velocity gradients make a complete turn from the disk to the envelope scales.

\noindent

%\textcolor{DarkMagenta}{Etant donné que le modèle présente initialement des fluctutations de densité mais aucune vitesse, nous avons lancé des simulations présentant ces mêmes fluctutations de densité, mais présentant aussi une rotation (donner beta), pour voir si les asymétries du modèle suffisent à désaligner les gradients petites et grandes échelles --- figures gradients avec rotation?. Les gradients ne sont pas parfaitement alignés dans ce cas non plus, mais leur amplitude de variation est beaucoup plus restreinte que dans le cas où le disque est issu du collapse asymétrique (et pas de la rotation initiale).}

%\noindent
%\begin{figure}
%\begin{center}
%\resizebox{\linewidth}{!}{\includegraphics{Figures/Angle_relatif_gradient_vitesse_B3%35_noturb_rot1_hydro_pert_asym_aleatoire50_vhr_output14_Rmin50_Rmax10000_dr50}}
%\end{center}
%\caption{Analyse de la simulation \emph{rot1}. Sont représentés les angles des %gradients de vitesse pour les 3 projections principales de la simulation. L'origine %est prise comme le gradient à très petite échelle, dans le disque. Temps physique de %la simulation=? A VOIR SUIVANT LE TEMPS DE LA SIMU SI LES GRADIENTS SONT AUTANT %DESALIGNES?}
%\label{angle relat rot1}
%\end{figure}

The amplitude of the velocity gradient at different scales is showed on the mid panel of Fig. \ref{gradients 50shr}. In the three projections the amplitude profile from a hundred astronomical units roughly follow a power law $j \propto R^{-1.8}$. It shows that these gradients are detectable in real observations, as the choice of appropriate molecular lines allows to detect gradient amplitude about \SI{1}{km.s^{-1}.pc^{-1}} in a solar type star-forming core at a distance of \SI{200}{pc}.%, \textcolor{DarkMagenta}{typical of Gould Belt regions.} %We also find that at scales larger than the hundred astronomical units away from the disk, the radial velocity is largely dominating the orthoradial one, meaning that this profile is due to the infall of the gas.

Once we have computed the amplitude of these gradients, we can determine the specific angular momentum. The evolution of this quantity at different scales is presented on the bottom panel of Fig. \ref{gradients 50shr}. This figure shows that for the three main lines of sight of the simulation, the specific angular momentum do not vary so much through the different scales, excepted some peaks correlated to abrupt changes in angular direction of velocity gradients. Furthermore, this quantity is roughly constant in the envelope, at the scale of the hundred and thousands astronomical units, with a mean value of about \SI{3e-4}{km.s^{-1}.pc}. In the discussion section, we compare the projected kinematics properties of our modeled core to observations in protostellar envelopes.
%This value is compatible with the results of \citet{Belloche2013} and \textcolor{DarkMagenta}{Gaudel et al. (2019)}.

\begin{figure}
\begin{center}
\resizebox{\linewidth}{!}{\includegraphics{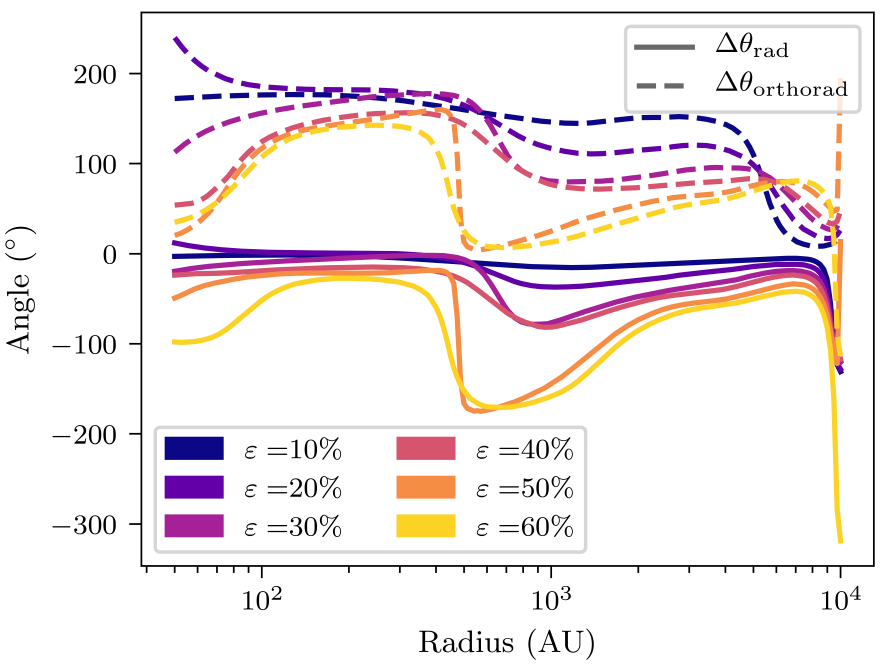}}
\end{center}
\caption{Angle between radial and total velocity gradients ($\Delta \theta_{\text{rad}}$, solid lines) and between orthoradial and total velocity gradients ($\Delta \theta_{\text{orthorad}}$, dashed lines), for different levels of perturbation $\varepsilon$. For the simulation with $10\%$ and $20\%$ of perturbations the total gradients directions follow the ones of the radial velocity gradients over all scales. For the simulations with more perturbations, $\Delta \theta_{\text{rad}}$ is higher from \SI{400}{AU} and $\Delta \theta_{\text{orthorad}}$ gets lower.}
\label{grad_rad_orthorad}
\end{figure}

To determine the origin of these velocity gradients, we compute the maps of projected velocities along the line of sight taking only the radial part of the velocity with respect to the disk center on the one hand, and the orthoradial part of the velocity on the other hand. We compute the velocity gradients directions over the different scales for the radial and orthoradial part of the velocity as we did for the total velocity gradients. We compared the angular deviation between the total velocity gradient and the radial velocity gradient --- $\Delta \theta_{\text{rad}}$ ---, and between the total velocity gradient and the orthoradial velocity gradient --- $\Delta \theta_{\text{orthorad}}$. The results for different levels of perturbation taken at similar times are visible Fig. \ref{grad_rad_orthorad}. For the simulation with $10\%$ of perturbations, the total and radial velocity gradient directions are very close over the probed scales\footnote{The smallest scale probed is \SI{50}{AU}, which is a bit more than the disk radius at the moment gradients are computed.}. As the level of perturbation increases, the radial and total gradients begins to misalign --- $\left| \Delta \theta_{\text{rad}} \right|$ increase --- for scales larger than \SI{400}{AU}, whereas the orthoradial and total gradients directions gets closer --- $\left| \Delta \theta_{\text{orthorad}} \right|$ decrease. These results depend on the line of sight choosen to compute the projected velocity, and to a lesser extent on the evolutionary stage of the simulations. What is a common feature is that for the small perturbation levels, the radial and total velocity gradients directions are very close over all the probed scales.

\noindent

\subsection{Size of formed disks}
\label{simu_sink}

\begin{figure*}
\begin{center}
\resizebox{0.5\linewidth}{!}{\includegraphics[width=\textwidth]{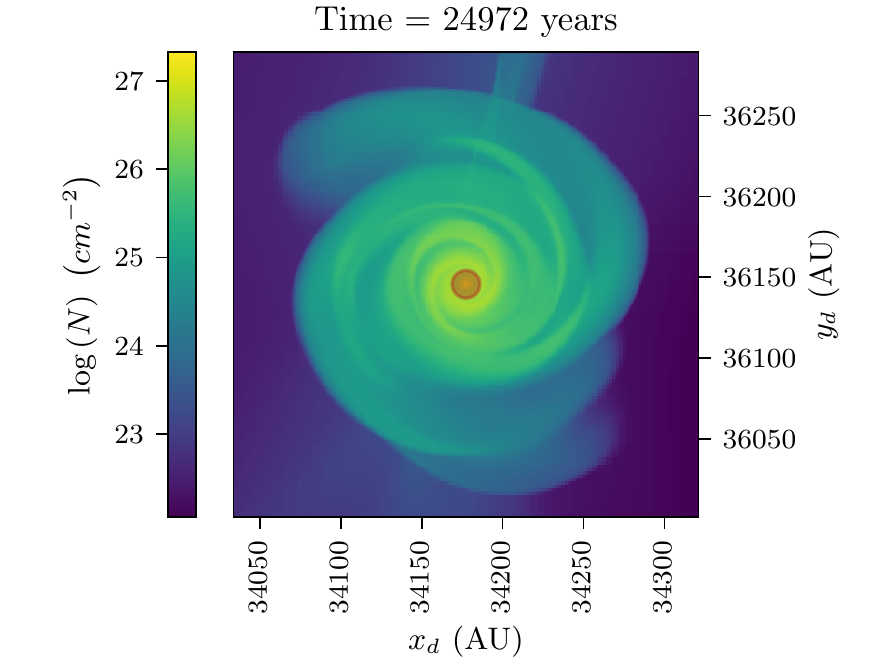}}%
\resizebox{0.5\linewidth}{!}{\includegraphics[width=\textwidth]{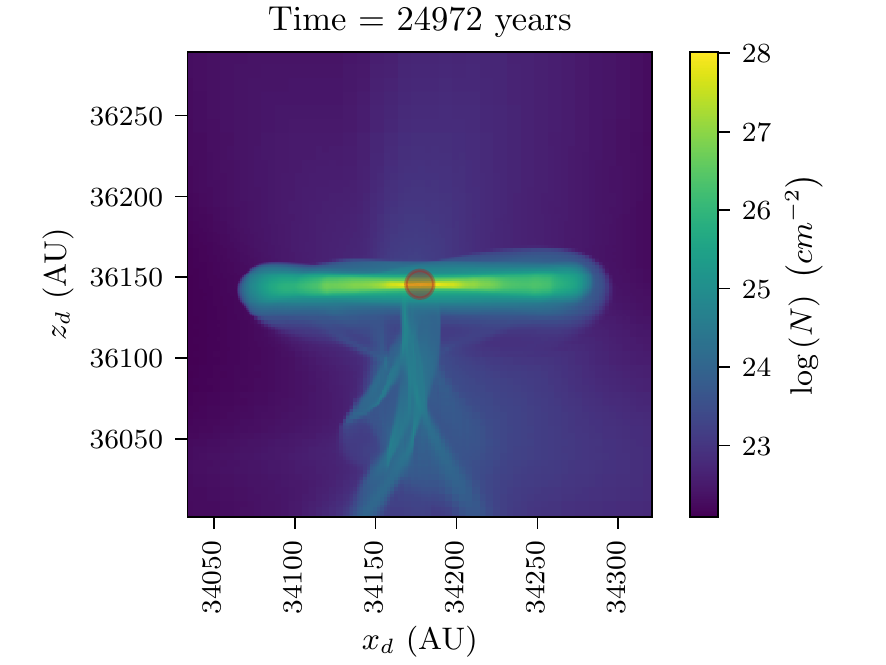}}

\caption{Column density in the simulation with $50 \%$ of perturbations and with sink particle (red circle). \emph{Left:} face-on projection. \emph{Right:} edge-on projection. At this time, the sink particle has a mass of \SI{0.64}{\Msun}. The disk is large with a radius of nearly \SI{125}{AU}.}%We see that a disk forms and grows in spite of the fact that initially the angular momentum is null.}
\label{disk_sink}
\end{center}
\end{figure*}

\begin{figure}
\begin{center}
\resizebox{\linewidth}{!}{\includegraphics[width=\textwidth]{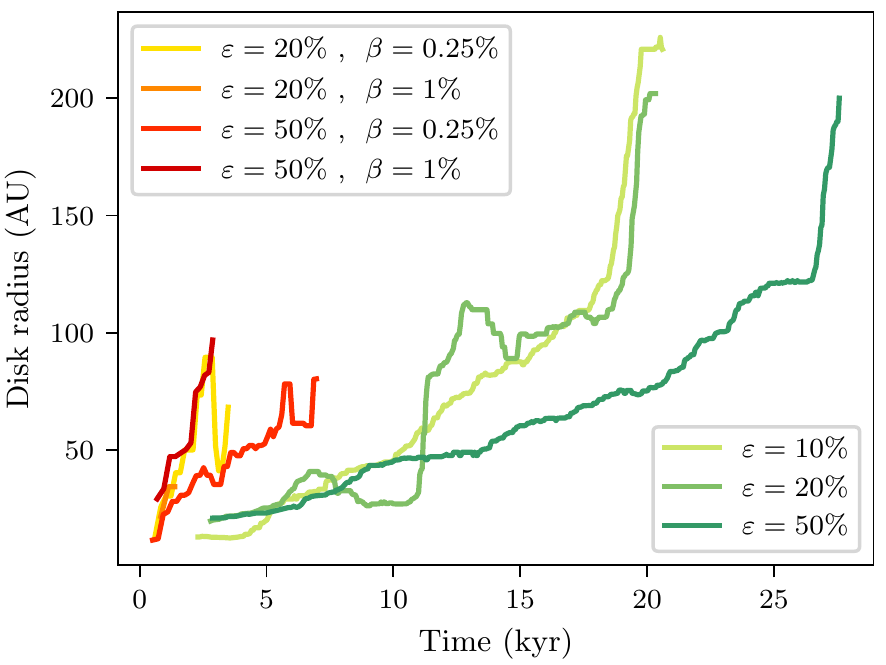}}%

\caption{Temporal evolution of disk size for a set of simulations. The three green curves are purely hydrodynamics simulations with different levels of perturbations $\varepsilon$. The four curves from red to yellow are also purely hydrodynamics, but include an initial solid-body rotation velocity profile, with different levels of perturbations and rotation $\beta$. The curves stop when the disk fragments. All the plotted curves have been smoothed by a sliding median.}%We see that a disk forms and grows in spite of the fact that initially the angular momentum is null.}
\label{taille_disk}
\end{center}
\end{figure}

We did not introduce until now any sink particle in our simulations. As a consequence, the gas cannot collapses at a smaller scale than our maximal resolution. This compels the gas to accumulate in the center of the disk, causing the disk to become autogravitating and leading to the apparition of the spiral arms visible in Fig. \ref{sequence disk} to transport angular momentum inside the disk. As this accumulation of dense gas in the disk is not physical --- the gas should continue to collapse to the stellar scales --- it is not correct to compute disk radius in these simulations. To handle this issue, we ran simulations with sink particles introduced when density reaches the threshold of \SI{e14}{cm^{-3}}. The disk grows but is less massive than previously, and match very well the keplerian profile. At some point the disk fragments. Figure \ref{disk_sink} shows the disk at an advanced stage, shortly before fragmentation. The disk reaches radius of about \SI{125}{AU}, which is large in comparison to observed disk radii \citep{Maury2010,Tobin2015,Segura-Cox2016,Maury2019}. To compute the disk radius in our simulations, the first step is to isolate what belongs to the disk. This selection is based on density and velocity criterion: a cell has to be dense enough and the orthoradial component of its velocity has to be at least two times larger than the radial one. Once this selection done, we look at the maximum distance projected in the equatorial plane for a large number of angular sectors. The mean of these distances over the different angular sectors is taken as the disk radius.
%The more straightforward method here would be to take the mean or the median of all these distances as the disk radius. As we will see in part \ref{part_MHD}, this method does not give coherent results for MHD simulations. Instead, we choose to construct the histogram of these distances, to take its peak, and finally to take the median of the distances in a range of \SI{16}{AU} centered on the peak value as the radius of the disk. The explanation of the necessity of this method is explained in part \ref{part_MHD}. 
Figure \ref{taille_disk} shows the evolution of disk radius over time for several simulations. The three green curves show the radius of the disk in the simulations with $10$, $20$ and $50 \%$ of perturbation level. It appears that in these three simulations the disks grow to reach radii around \SI{200}{AU} in $20$ to \SI{30}{kyr}, before fragmenting. The disk size do not depends much on the perturbation level in the range of $10$ to $50\%$ of perturbations.

\noindent

%\noindent
%\begin{figure*}
%\begin{center}
%\resizebox{0.5\linewidth}{!}{\includegraphics[width=\textwidth]{Figures/Taille_disque_all_modif.pdf}}%
%\resizebox{0.5\linewidth}{!}{\includegraphics[width=\textwidth]{Figures/Taille_disque_zoom.pdf}}

%\caption{Disk size versus time for a set of simulations. The three green curves are purely hydrodynamical simulations with different level of perturbations $\varepsilon$. The four curves from red to yellow are similar simulations, but include an initial solid-body rotation velocity profile, with different level of perturbations and rotation $\beta$. The light blue and purple curves correspond to MHD simulations, including or not initial rotation. The right panel is a zoom corresponding to the blue rectangle in the left panel. For the seven hydrodynamical simulations, the curves stop when the disk fragments. The two MHD simulations do not fragment during the simulation. All the plotted curves have been smoothed by a sliding median.}%We see that a disk forms and grows in spite of the fact that initially the angular momentum is null.}
%\label{taille_disk}
%\end{center}
%\end{figure*}

\noindent

\subsection{Effects of initial rotation}
\label{part_rotation}
To see the influence of initial rotation on our results, we ran four simulations in which a solid-body rotation velocity profile is imprinted in the initial conditions. We ran simulations with $\varepsilon = 20 \%$ and $50 \%$, and for each of these perturbation levels we choose two rotation levels\footnote{$\beta$ is the ratio of rotational over gravitational energy.} $\beta = 0.25 \%$ and $1 \%$. The evolution of the disk radius in these simulations is visible in Fig. \ref{taille_disk}, represented by the four curves from red to yellow. For these four simulations the evolution of the radius is similar. These disk grows rapidly before fragmenting between $2$ to \SI{7}{kyr} at a radius around \SI{100}{AU}. This evolution is different from the ones of the disks in the simulation without initial rotation, in particular because the disk forms earlier and grows faster. While when there is no rotation, it takes about \SI{15}{kyr} to get a disk bigger than \SI{100}{AU}, this takes 
only 3 to \SI{5}{kyr} when rotation is included.

\begin{figure}
\begin{center}
\resizebox{\linewidth}{!}{\includegraphics{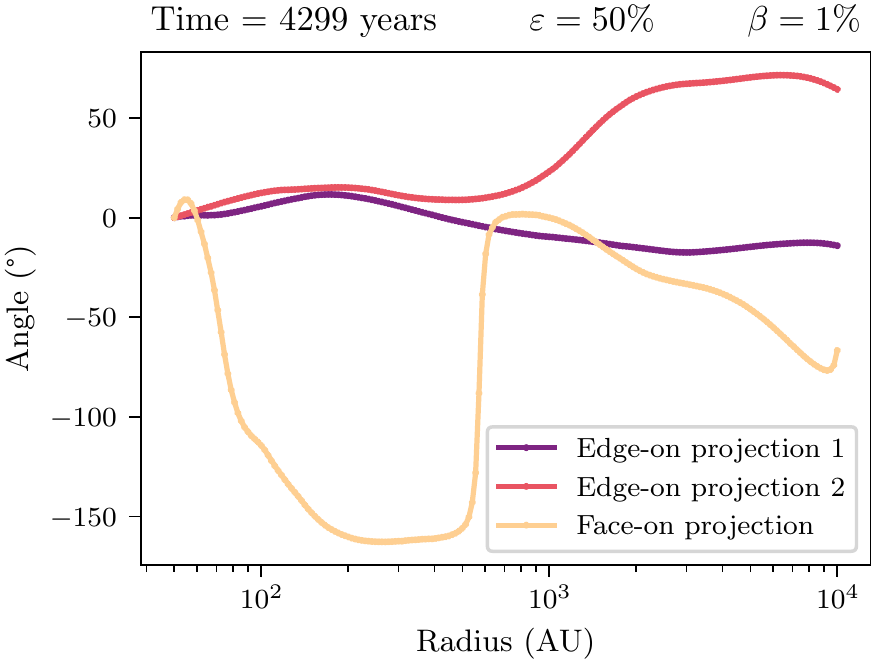}}

\resizebox{\linewidth}{!}{\includegraphics[trim = 0cm 0cm 0cm 0.3cm, clip]{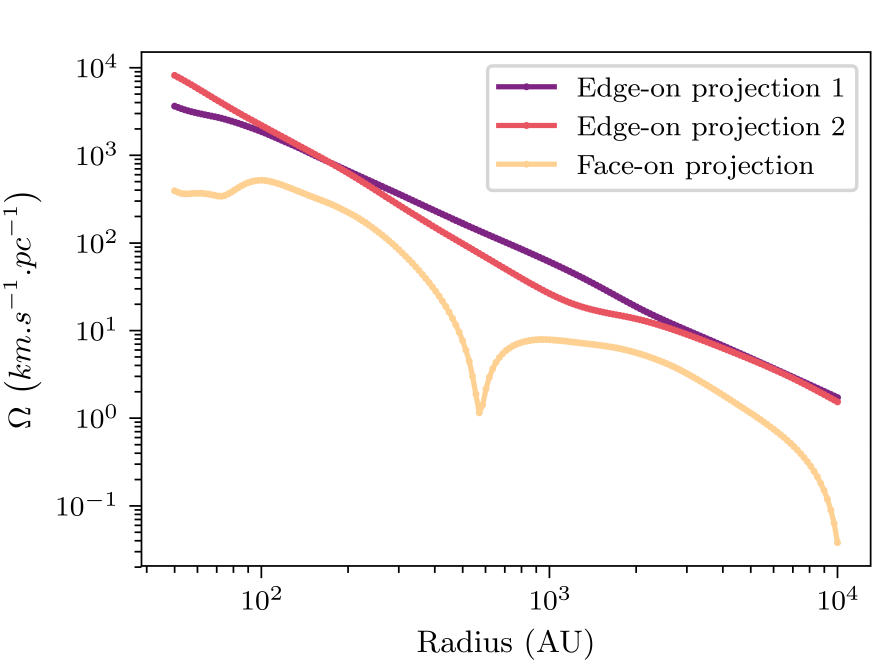}}

\resizebox{\linewidth}{!}{\includegraphics[trim = 0cm 0cm 0cm 0.3cm, clip]{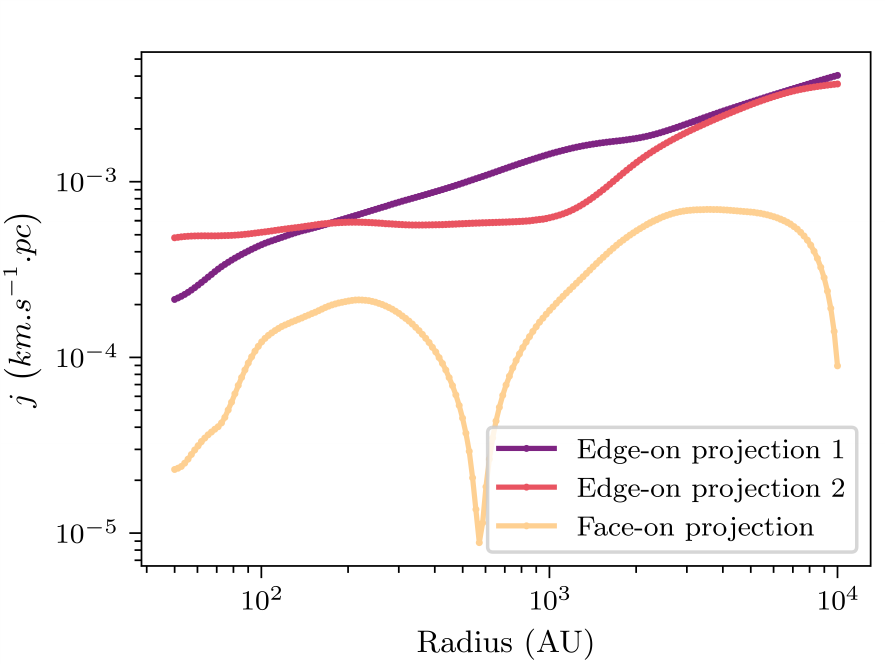}}
\end{center}
\caption{Analysis of velocity gradients at different scales in the simulation with $50\%$ of perturbations and $\beta = 1\%$. On each panel, the three curves correspond to two edge-on and one face-on projections. \emph{Top}: angular direction of velocity gradients. The origin of the angular direction corresponds to the direction of the disk scale gradient. \emph{Middle}: amplitude of velocity gradients. \emph{Bottom}: specific angular momentum as computed in observation analysis.}
\label{velocity_grad_with_rot}
\end{figure}

\begin{figure}
\begin{center}
\resizebox{\linewidth}{!}{\includegraphics{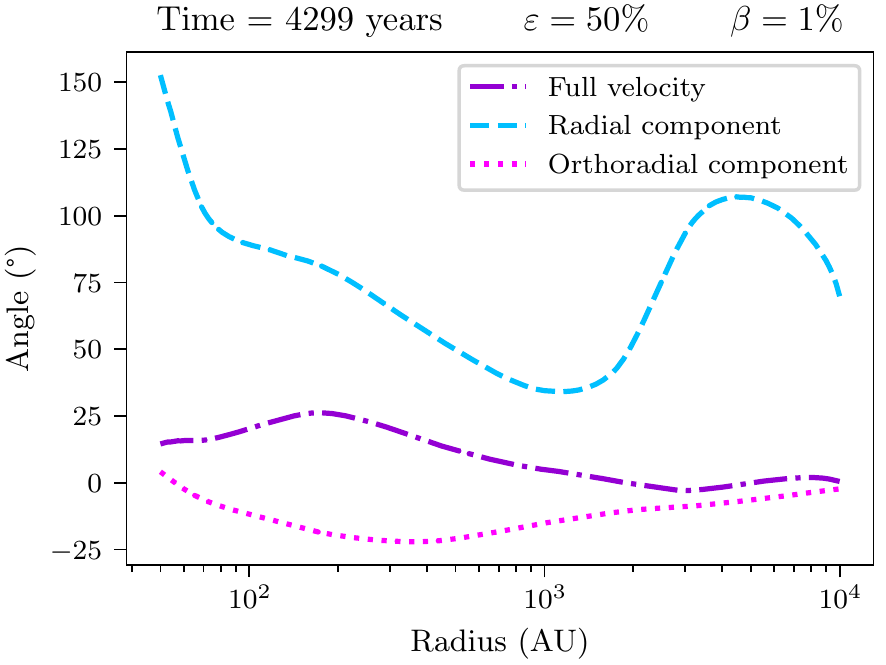}}

\resizebox{\linewidth}{!}{\includegraphics[trim = 0cm 0cm 0cm 0.3cm, clip]{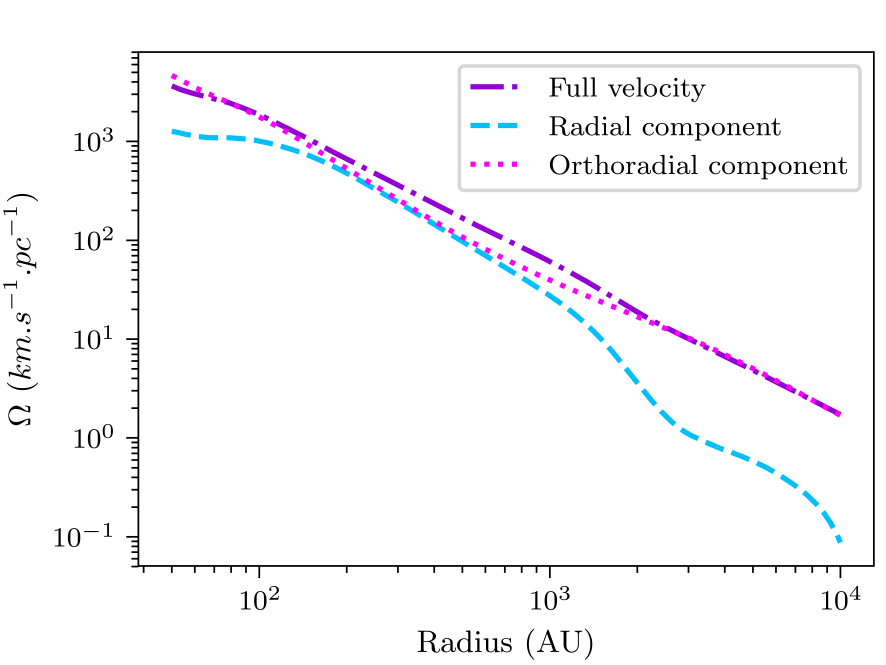}}

\resizebox{\linewidth}{!}{\includegraphics[trim = 0cm 0cm 0cm 0.3cm, clip]{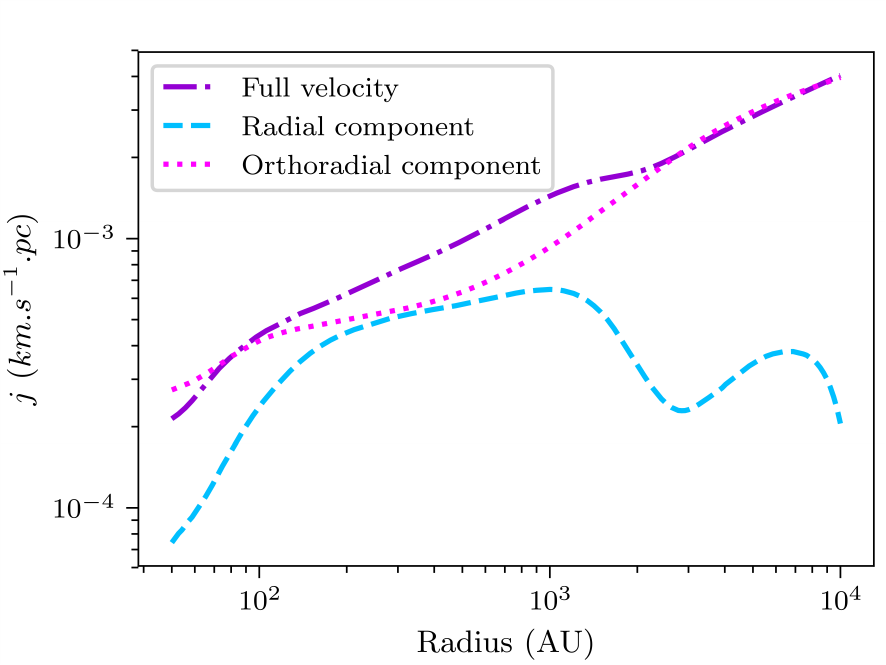}}
\end{center}
\caption{Analysis of velocity gradients for the full velocity (in purple dashdotted), the radial component (in blue dashed), and the orthoradial component (pink dotted). These quantity are represented at different scales in the simulation with $50\%$ of perturbations and $\beta = 1\%$, for the edge-on projection 1. \emph{Top}: angular direction of velocity gradients. The origin of the angular direction corresponds to the direction of the disk scale gradient. \emph{Middle}: amplitude of velocity gradients. \emph{Bottom}: specific angular momentum as computed in observation analysis.}
\label{grad_rad_ortho_rot}
\end{figure}

We conducted in these simulations the same analysis of velocity gradients than in part \ref{part_velocity_gradients}. The results for the simulation with $\varepsilon = 50\%$ and $\beta = 1 \%$ are shown in Fig. \ref{velocity_grad_with_rot} and \ref{grad_rad_ortho_rot}. In the top panel of Fig. \ref{velocity_grad_with_rot}, we see that the angular deviations of the velocity gradients are less important than in the case without initial rotation. Only the face-on projection exhibit large deviations, but the amplitude $\Omega$ in this projection is much smaller than in the two edge-on projections, by a factor around $10$. The resulting specific angular momentum is visible on the bottom panel of Fig. \ref{velocity_grad_with_rot}. For the two edge-on projections, it exhibit values that are larger than in the case without initial rotation, and that are larger than the ones deducted from observations. We also see that  $j \propto R^\alpha$ with $\alpha \simeq 0.5$ while 
observations revealed that in the inner few thousands
AU, $\alpha \simeq 0$ (see \citep{Belloche2013} for a review). The most recent observations by \citet{Gaudel2020} reveal $\alpha \simeq 0.3 \pm 0.3$  under \SI{1600}{AU}.

Here also we computed the decomposition of the velocity in its radial and orthoradial components and conducted the same analysis of velocity gradients with these two components. The results are visible in Fig. \ref{grad_rad_ortho_rot} for the full velocity gradients, the radial velocity gradients and and orthoradial velocity gradients, in the simulation with $\varepsilon=50\%$ and $\beta=1\%$, for the edge-on projection 1. The top panel shows the angle using the same arbitrary reference for the three gradients, whereas the middle and bottom panels show respectively the amplitude and the specific angular momentum deducted from these gradients. The joint analysis of the top and middle panel shows that for $R<\SI{e2}{AU}$ and $R>\SI{3e3}{AU}$ the orthoradial velocity gradient is dominant, as it is much larger in amplitude than the radial one and much closer to the full velocity gradient in angular position than the radial one is. This means that at large scale the full velocity gradient is tracing the rotation of the envelope. For $\SI{e2}{AU} < R < \SI{3e3}{AU}$, the contribution of the radial and orhoradial component are of the same order of magnitude in amplitude, which leads to the full velocity gradients to be misaligned with both the radial and orthoradial gradients at those scales. In the face-on projection the effect of the rotation is nearly invisible and the results are similar to the ones presented part \ref{part_velocity_gradients}.

The results are comparable for the four simulations introduced in this part and therefore the other cases are not shown for conciseness.

\noindent

\noindent

\subsection{Effects of the magnetic field}
\label{part_MHD}

\begin{figure*}
\begin{center}
\resizebox{0.5\linewidth}{!}{\includegraphics[width=\textwidth]{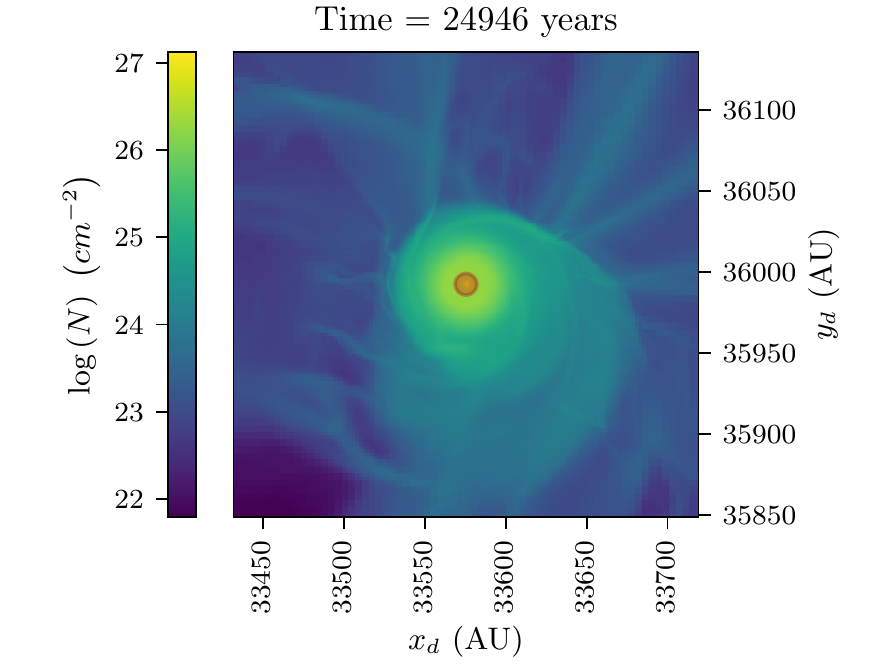}}%
\resizebox{0.5\linewidth}{!}{\includegraphics[width=\textwidth]{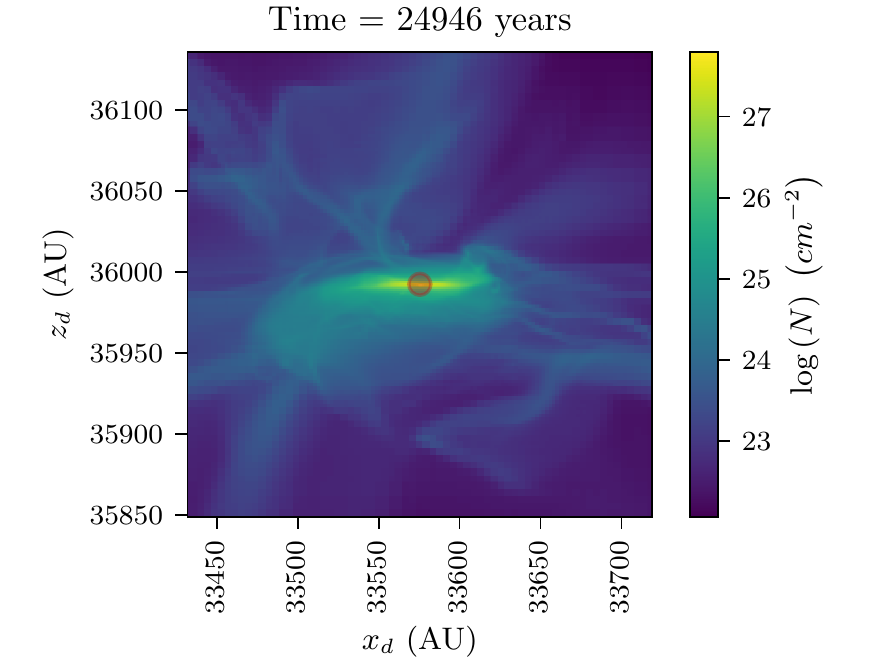}}

\caption{Column density in the MHD simulation with $50 \%$ of perturbations and with sink particle (red circle), at the same time than Fig. \ref{disk_sink}. \emph{Left:} face-on projection. \emph{Right:} edge-on projection. At this time, the sink particle has a mass of \SI{0.40}{\Msun}. It is hard to define a proper disk radius in this case.}%We see that a disk forms and grows in spite of the fact that initially the angular momentum is null.}
\label{fig_MHD}
\end{center}
\end{figure*}

For the sake of completeness, we finally add magnetic field\footnote{We set a magnetisation $\mu = 0.3$. For the MHD simulation with initial rotation, the axis of the magnetic field and the axis of rotation are aligned.} in our simulations to see its effects on the formed disk. The magnetic field is treated under the ideal MHD approximation. In these simulations a disk still forms even in the absence of initial large scale rotation. Figure \ref{fig_MHD} shows the appearance of the disk at the same time than Fig. \ref{disk_sink}. Clearly the two disks are qualitatively very different. In the purely hydrodynamics case, the disk is big and massive, with sharp edges, whereas in the MHD case, the disk is smaller and less massive (smaller column density).

In the MHD case it is hard to define properly a disk radius. In fact, when looking at Fig. \ref{fig_MHD}, the disk is buried in a very filamentary structure that has the same order of magnitude in column density as the outer part of the disk. At some places these filaments verify the velocity criteria set out in part \ref{simu_sink} to define the belonging to the disk. Here we are confronted to a definition problem. In hydrodynamics simulations, the disk has sharp edges and it is clear to define by eye what belongs to the disk or not. In MHD simulations it is really hard to do so and therefore we do not attempt 
here to present a quantitative comparison.
%In Fig. \ref{fig_MHD}, should we consider as disk only the inner dense part of around \SI{25}{AU} in radius, or should we extend it to \SI{56}{AU} as found by our algorithm? This shows that our algorithm largely overestimates the disk radius for MHD simulations.

%Being aware of this issue, we can compare by eye 
In spite of this difficulty, the comparison between
Figs. \ref{disk_sink} and \ref{fig_MHD} reveals that in the MHD case, the disk is much smaller. For the MHD simulation without initial rotation, the disk stops growing after \SI{20}{kyr} and does not fragment like in the hydrodynamics case. For the simulation with initial rotation, the disk grows more rapidly but also stabilises at the same size between 40 to \SI{70}{AU} depending on the definition of what belongs to the disk.
We believe that this is the signature of the magnetic braking occurring in magnetised disk.

\noindent

\section{Discussion}

From a very simple model, we showed that the angular momentum computed in the frame of the disk in relation to the center of the disk is not a conserved quantity. This is due to the non Galilean nature of the frame, provoked by the non-axisymmetrical gravitational collapse. This collapse leads to the formation of a protostellar disk. 
%In our system, the current paradigm --- the inheritance of large scale rotation through angular momentum conservation --- is incorrect. 
In our system, the angular momentum that forms the disk is not a mere conservation of pre-existing large scale angular momentum. 
The rotation can be generated locally by the asymmetry of the collapse. This result is in agreement with the early work by \citet{Peebles69}. This mechanism responsible for the formation of spiral galaxies is thus also able to account for the formation of protostellar disks. Thereby, the conservation of a pre-existing angular momentum at large scales might not be the only mechanism responsible for protostellar disk formation, and the assumption that small scale rotation is being inherited from the large scales may not be correct, at least for some systems. 

The disks formed in the simulations with sink particles reach radius larger than a hundred astronomical units until they fragment. As these simulations start with no angular momentum, but only require moderate density perturbations --- which are expected to be present in real cores ---, this study gives a sort of ``minimal hydrodynamic radius'' that a disk should reach in a hydrodynamics collapse. As most of the observed disks are much smaller \citep{Maury2010,Tobin2015,Segura-Cox2016,Maury2019}, it implies that angular momentum extraction processes are at work in real disks. 
%With simple axisymmetrical hydrodynamic collapse models in which rotation is initially present, the size of the disk formed depends on the initial rotation. One can ask themselves if the small sizes of the disks obtained from real observations are due to a low level of rotation, or if processes extracting angular momentum are at work. The fact that those processes should exist is well accepted \textcolor{DarkMagenta}{(réf?)}, so this result is not new, but here we presented a new argument in favour of these processes. 
In the traditional picture of axisymmetric collapse, the size of the disk directly depends on the initial rotation. The presence of small disks in observations could therefore be interpreted as due to low rotation levels. In the present study, we show that in the non-axisymmetric case, even modest levels of density perturbations lead to the formation of large hundred of AU size disks. These latter appear to be natural outcome of the gravitational dynamics and local conservation of angular momentum. These are generic features. Therefore the most natural way to reconcile these observations with the generic nature of big hydrodynamical disks is to invoke the presence of magnetic braking that we know is operating in cores \citep{Li2014,Hennebelle2019}. Our MHD simulation starting from the same initial conditions but just adding a magnetic field shows that the outcoming disk is much smaller and do not fragments, as expected from these previous studies.

The magnitudes of the velocity gradients from our modeled core do not depend much on the chosen projection, although a dip is observed in the $z$ projection at scales of \SI{600}{AU} (see mid panel of Fig. \ref{gradients 50shr}) which results from a reversal in angular direction of velocity gradients. Their amplitude is roughly consistent with velocity gradients observed in protostellar cores using dense gas tracers (see for example \citet{Chen2007,Belloche2004} who recover amplitudes from \SI{0.1}{} to \SI{10}{km.s^{-1}.pc^{-1}} in protostellar cores at scales of \SI{5000}{AU}), but also with the recent results of \citet{Gaudel2020} probing inner regions of Class 0 envelopes, down to a hundred AU.
The specific angular momentum $j$ in our modeled core --- computed from the velocity gradients following a similar methodology as the one extracting $j$ from observations --- has a roughly constant value of about \SI{3e-4}{km.s^{-1}.pc} between \SI{e2}{AU} and \SI{e3}{AU} (see bottom panel of Fig. \ref{gradients 50shr}).
This order of magnitude is similar to typical specific angular momentum values recovered from observations in solar-type protostellar cores: see \citet{Yen2015a} for observations in 7 Class 0 protostars and the work of \citet{Gaudel2020} in 12 Class 0 protostars giving a specific angular momentum of \SI{5e-4}{km.s^{-1}.pc} below \SI{1400}{AU}.
The few observational constraints on the spatial distribution of specific angular momentum in prestellar cores and protostars suggest a scenario where local specific angular momentum is following a power-law $R^{1.6}$ in starless structures at scales larger than \SI{6000}{AU}, while it is constant and conserved within collapsing star-forming cores (see \citet{Belloche2013} for a review). However, \citet{Yen2015b} found that the decreasing trend of $j(R)$ observed at large scales propagates down to radii smaller than \SI{5000}{AU}. \citet{Gaudel2020} resolved the break radius around \SI{1600}{AU} where it stabilizes with a weak dependence on radius $j \propto R^{0.3 \pm 0.3}$ between $50$ and \SI{1600}{AU}. These observations are very well matched by the properties of specific angular momentum in our simulations with density perturbations and no rotation. Our model shows a dependency of $\Omega$ along a $R^{-1.8}$ power-law, which translates ($j=R^2 \Omega$) as a $j  \propto R^{0.2}$ power-law, centered on the value of \SI{3e-4}{km.s^{-1}.pc}, which coincides with the recent observational results cited earlier.
%We stress that, in our model, the angular momentum at scales larger than \SI{100}{AU} is created by the infall motions of the core, which nature cannot be distinguished from specific angular momentum which would be due to rotation when examining only the observed specific angular momentum values.

However, the analysis of velocity gradients also reveals a large dispersion of the gradients directions between the disk and envelope scales. Depending on the viewing angle, the observed velocity gradients in the envelope can be very misaligned with respect to the small-scale gradient resulting from disk rotation. The projected kinematics can even produce a complete reversal of the gradient at some scales: for example in top panel of Fig. \ref{gradients 50shr}, the $y$ projection (red curve) shows a gradient about \SI{200}{\degree} from the disk scale gradient at scales \SI{2000}{AU}. Since the rotation is generated locally in our model, it is not surprising that small and large scales gradients are unrelated and hence misaligned. 
In our model there is no large scale rotation, suggesting that observations of strongly misaligned gradients in protostellar cores could be due to non-axisymmetric gravitational collapse, rather than to global rotational motions of the core.
%In our model, velocity gradients at large scales are mostly due to infalling motions rather than rotation, suggesting that observations of strongly misaligned gradients in protostellar cores could be due to contamination by asymetric infall motions at the large scales, rather than to global rotational motions of the core. 
Such scenario could explain, for example, the reversal of velocity gradient observed in the L1527 protostar \citep{Tobin2011}, or more recently our findings that the rotation of the small-scale disk in the low-luminosity protostar IRAM04191 (Maury et al. in prep) is opposite to the large-scale gradient at \SI{2000}{AU} scales in the envelope, interpreted as core rotation \citep{Belloche2004}. This observation of a ``counter-rotating'' disk is incompatible with simple axisymmetrical collapse models in which the disk would have simply formed because of the conservation of a large scale angular momentum in a rotating core. \citet{Tsukamoto2017} developed MHD models including Hall effect that form counter-rotating envelopes at the upper region of a pseudo-disk under some conditions about the alignment between magnetic field and initial rotation. This thin counter-rotating layer is located close to the pseudo-disk, typically between $50$ and \SI{200}{AU}, and are thus unable to explain counter-rotation in the outer envelopes, at scales larger than  \SI{1000}{AU}, while it is a natural feature of our model.

The analysis of the radial and orthoradial part of the velocity and their contribution to the total velocity gradient directions shows that for small perturbation levels (typically less than 20\%), the radial and total velocity gradients directions are very close over all the probed scales whereas the orthoradial component is misaligned. As the perturbation level increase, for scales larger than \SI{400}{AU} the radial component is more and more misaligned whereas the orthoradial one tends to be closer to the total velocity gradient direction. These features depends on the choice of the line of sight. A common point which seems independent from the choice of a line of sight is that for small perturbation levels, the radial component is very close to the total velocity gradient direction. As the simulations start with no global rotation, this analysis shows that the notion of \emph{radial} and \emph{orthoradial} though as infall and rotation is less and less clear as the level of perturbations increases, which means as the deviation from axisymmetry grows. With a small perturbation level, the deviation from axisymmetry is low and the velocity gradients are mainly due to infall motions. With large perturbation levels, the deviation from axisymmetry is high and the velocity gradients are resulting from the complex dynamics of the self-gravitating gas.

When adding global rotation to the initial conditions, we found that the  formed disks grow more rapidly and fragment earlier. Analysing the kinematics of the gas in those simulation with initial rotation, we show that the angular deviations are lower than in the case without initial rotation, which is less in agreement with \citet{Gaudel2020} observational results. From this analysis, the inferred specific angular momentum exhibits a slope and a mean value larger than in the case without initial rotation. As for the angular deviation, this result is less in agreement with observations than the case without initial rotation.

\section{Conclusion}

We have shown that protostellar disks can emerge from a non-axisymmetrical gravitational collapse in
which there is no rotation initially. 
We ran purely hydrodynamical simulations of a collapsing dense core starting from an initial condition where all the cells are at rest, and we broke the symmetry of the problem by adding density fluctuations over a flat profile. We showed analytically that the angular momentum in the frame of the accretion center and with respect to the accretion center is not a conserved quantity due to the non Galilean character of the frame. This leads to the possibility of the formation of a disk 
and we demonstrated numerically that a disk indeed forms in these simulations. 

We then analysed our simulations from an observational point of view,  computing the velocity gradients at different scales as it is done with real observations, and deducing the amplitude of these gradients and the specific angular momentum over the different scales. The results we obtained for the value of the specific angular momentum matches the ones from \citet{Belloche2013} and \citet{Gaudel2020}.

This study then suggests a new paradigm for the formation of protostellar disks but does not replace the current one --- the conservation of pre-existing angular momentum at large scale. It shows that even if this conservation of angular momentum is absolutely correct in axisymmetrical model, it is more complicated when the axisymmetry is broken, as the accretion center frame becomes non Galilean. In the extreme case we studied in which every cells are initially at rest, it shows that even without initial rotation the formation of a protostellar disk is possible. The formation of these disks in our model does no longer depends on specificities of large scales, but results in a more generic process only due to the density fluctuations of the gas. We showed that the different features of the model based on the analysis of velocity gradients are realistic. This model could thus help understand the features observed in some objects, as the angular deviation of velocity gradients at different scales. It also helps to understand the formation of a disk within progenitors that does not seems to contain enough angular momentum at large scales to form a disk by conservation of the angular momentum during the collapse. We found that these models without initial rotation are in better agreement with observational results than the models with initial solid-body rotation.

Our model uses minimal physical ingredients: hydrodynamics, thermodynamics, gravity, and density fluctuations. It is thus robust in the sense that all these ingredients are always present in dense cores. In particular, we found that even for $10 \%$ of density perturbations initially, a disk forms.  Protostellar disks appear to be natural features of a gravitational collapse as soon as it is not axisymmetric. As large disks are formed in this new scenario, we stress the necessity to evoke processes extracting angular momentum to lead to the formation of smaller disks, as found with observations. 
%But our model is also limited by the fact that it does not include several important physical processes. 
We showed that the presence of magnetic field reduces the size of the formed disk, in agreement with past studies \citep{Hennebelle2016,Maury2018}.
\begin{acknowledgements}
We sincerely acknowledge the anonymous referee for the revision of our work and it's comments and suggestions.
\end{acknowledgements}
%============================================================================

\bibliography{References}

%============================================================================
\begin{appendix}
\section{Full theoretical development}
\label{annexe dev theorique}
In this annex we reused the notations and definitions of part \ref{formation disque theorique}.

\subsection{Development from the angular momentum in the box frame}
\label{dev_theor_part_1}
The frame $\mathcal{R}$ of the simulation box is Galilean and there is not any external forces that applied on the system of material points. The angular momentum $\left. \bm{\sigma_O} \right| _{_{\mathcal{R}}}$ computed in $\mathcal{R}$ in relation to $O$ is thus conserved. We will consider for the entire development an initial condition where all the points are motionless in $\mathcal{R}$, this ensure that $\left. \bm{\sigma_O} \right| _{_{\mathcal{R}}}$ is equal to zero and stay null through the temporal evolution of the system.

\begin{equation}
\left. \bm{\sigma_O} \right| _{_{\mathcal{R}}} = \sum_i m_i \bm{OM_i} \wedge \frac{\d \bm{OM_i}}{\d t} = \bm{0}
\end{equation}

\noindent
From this expression, we can involve the accretion center $C$:

\begin{equation}
\begin{split}
 \sum_i m_i \bm{OC} \wedge \frac{\d \bm{OC}}{\d t} + \sum_i m_i \bm{CM_i} \wedge \frac{\d \bm{OC}}{\d t} +  \\ \sum_i m_i \bm{OC} \wedge \frac{\d \bm{CM_i}}{\d t}  + \sum_i m_i \bm{CM_i} \wedge \frac{\d \bm{CM_i}}{\d t}  =  \bm{0} 
\end{split}
\end{equation}

\noindent
Involving the expression of the total mass $M = \sum\limits_i m_i$, using the definition of center of mass $G$ which implies $\sum\limits_i m_i \bm{CM_i} = M \bm{CG}$, and recognising\footnote{We choose $\mathcal{R'}$ to be in translation with respect to $\mathcal{R}$ in order to have the equality of the temporal derivative operators in the two frames, without loosing any generality.}  that the forth term of the expression above if the angular moment $\left. \bm{\sigma_C} \right| _{_{\mathcal{R'}}}$ defined Eq. (\ref{moment reel}), we obtain:

\begin{equation}
M \bm{OC} \wedge \frac{\d \bm{OC}}{\d t} + M \bm{CG} \wedge \frac{\d \bm{OC}}{\d t} + M \bm{OC} \wedge \frac{\d \bm{CG}}{\d t} + \left. \bm{\sigma_C} \right| _{_{\mathcal{R'}}} = \bm{0}
\end{equation}

\noindent
Rewriting the second term to make the point $G$ appear:

\begin{equation}
\begin{split}
M \bm{OC} \wedge \frac{\d \bm{OC}}{\d t} + M \bm{CG} \wedge \frac{\d \bm{OG}}{\d t} + M \bm{CG} \wedge \frac{\d \bm{GC}}{\d t} \\ + M \bm{OC} \wedge \frac{\d \bm{CG}}{\d t} + \left. \bm{\sigma_C} \right| _{_{\mathcal{R'}}} = \bm{0}
\end{split}
\end{equation}

\noindent
Reuniting the first term with the forth one, and then with the second one, we obtain:

\begin{equation}
M \bm{OG} \wedge \frac{\d \bm{OG}}{\d t} + M \bm{CG} \wedge \frac{\d \bm{GC}}{\d t} + \left. \bm{\sigma_C} \right| _{_{\mathcal{R'}}} = \bm{0}
\label{avant dernier dev}
\end{equation}

\noindent
We can applied the inertia center theorem to the whole system of material points in $\mathcal{R}$. In the lack of external force, it gives:

\[
M \frac{\d ^2 \bm{OG}}{\d t^2} = \bm{F_{\text{ext}}} = \bm{0} ~~~\Longrightarrow ~~~ \frac{\d \bm{OG}}{\d t} = \bm{\text{cte}} = \bm{0} 
\]

\noindent
because the velocity in each point is initially null. Then only two terms remain in the Eq. (\ref{avant dernier dev}), which can be rewritten in the form of the Eq. (\ref{eq moment dans R'}): 

\begin{equation}
\left. \bm{\sigma_C} \right| _{_{\mathcal{R'}}} = M \bm{GC} \wedge \frac{\d \bm{GC}}{\d t}
\label{eq mom fin}
\end{equation}

\noindent
This expression is equivalent to $\left. \bm{\sigma_C} \right| _{_{\mathcal{R'}}} = M \bm{GC} \wedge \frac{\d \bm{OC}}{\d t}$ due to the fact that $\frac{\d \bm{OG}}{\d t} = \bm{0}$.

\subsection{Interpretation with the inertial force}
\label{inertial_force_development}
Let's consider the problem from the frame $\mathcal{R'}$. The latter is not Galilean, and we choose it to be in translation with respect to $\mathcal{R}$, without loosing any generality, in order to have the equality of the temporal derivative operators in $\mathcal{R}$ and $\mathcal{R'}$. We can thus rewrite the evolution equation of the angular momentum $\left. \bm{\sigma_C} \right| _{_{\mathcal{R'}}}$ in the absence of external force:

\begin{equation}
\frac{\d \left. \bm{\sigma_C} \right| _{_{\mathcal{R'}}}}{\d t} = \sum_i \bm{\mathcal{M}_C} \left( \bm{F}_{\text{ie} \rightarrow i} \right)
\label{eq Fie}
\end{equation}

\noindent
where $\bm{\mathcal{M}_C \left( \bm{F}_{\text{ie} \rightarrow i} \right) }$ is the torque of the inertial force that is exerted on each point $i$, in relation to the point $C$. For the given conditions of $\mathcal{R'}$, this force can be written:

\begin{equation}
\bm{F}_{\text{ie} \rightarrow i} = -m_i \frac{\d ^2 \bm{OC}}{\d t^2}
\end{equation}

\noindent
We can thus write the momentum of $\bm{F}_{\text{ie} \rightarrow i}$ in relation to the point $C$ as :

\begin{equation}
\bm{\mathcal{M}_C} \left( \bm{F}_{\text{ie} \rightarrow i} \right) = - m_i \bm{CM_i} \wedge \frac{\d ^2 \bm{OC}}{\d t^2}
\end{equation}

\noindent
Reusing the definition of the center of mass and the fact that $\frac{\d \bm{OG}}{\d t} = \bm{0}$, the right side of Eq. (\ref{eq Fie}) becomes :

\begin{equation}
\sum_i \bm{\mathcal{M}_C} \left( \bm{F}_{\text{ie} \rightarrow i} \right) = - \sum_i m_i \bm{CM_i} \wedge \frac{\d ^2 \bm{OC}}{\d t^2} = M \bm{GC} \wedge \frac{\d ^2 \bm{GC}}{\d t^2}
\end{equation}

\noindent
Eq. (\ref{eq Fie}) then becomes :
\begin{equation}
\frac{\d \left. \bm{\sigma_C} \right| _{_{\mathcal{R'}}}}{\d t} = M \bm{GC} \wedge \frac{\d ^2 \bm{GC}}{\d t^2}
\end{equation}

\noindent
This expression is the same that the one obtained by taking the derivative of Eq. (\ref{eq mom fin}), but allows to understand that the non conservation of the angular momentum $\left. \bm{\sigma_C} \right| _{_{\mathcal{R'}}}$ is due to the non-Galilean character of the frame \ensuremath{\mathcal{R'}}.

\newpage
\section{Numerical validity}
\label{annexe conv num}

In this annex we reused the notations of part \ref{analyse 50vhr}.

\begin{figure*}
\begin{center}
\resizebox{0.5\linewidth}{!}{\includegraphics[width=\textwidth]{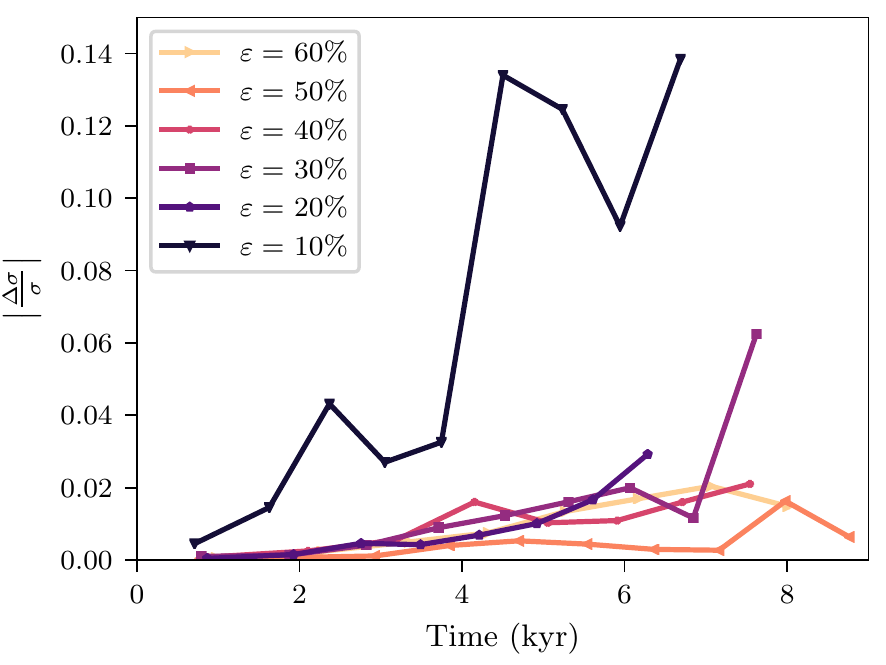}}%
\resizebox{0.5\linewidth}{!}{\includegraphics[width=\textwidth]{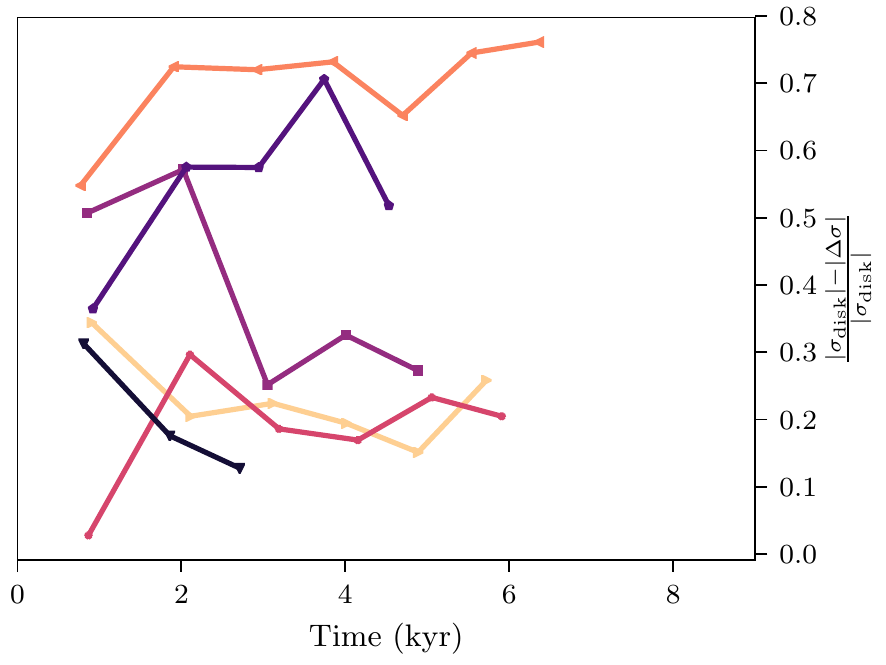}}

\caption{Analysis of numerical errors in the simulations with different levels of perturbations $\varepsilon$, without sink particle. The legend is the same for both panels. To improve visibility, the results have been smoothed. \emph{Left:} the relative difference between $\sigma_\text{num}$ and $\sigma_\text{an}$. \emph{Right:} the relative difference between the momentum in the disk $\left| \sigma_{\text{disk}} \right|$ and $\left| \Delta \sigma \right| = \left| \sigma_\text{num} - \sigma_\text{an} \right|$. We calculate the momentum in the disk only when the disk is not fragmented. The simulation with $10\%$ of perturbations shows non negligible amount of numerical errors, whereas all the other simulations can be trusted.}
\label{conv_num_fig_all}
\end{center}
\end{figure*}

We showed in part \ref{analyse 50vhr} that looking at the difference between $\sigma_\text{num}$ and $\sigma_\text{an}$ and comparing this difference to the momentum in the disk gives us a simple way to estimate the trust level of our simulations. The results for all the perturbation levels we studied are given in Fig. \ref{conv_num_fig_all}. The left panel shows the relative difference between $\sigma_\text{num}$ and $\sigma_\text{an}$. Except for simulation with $\varepsilon=10\%$, this difference is less than a few percent. For the simulations with $10\%$ of perturbation this difference grows higher.

In order to conclude on the trust level on the simulations, we have to compare $\left| \Delta \sigma \right| = \left| \sigma_\text{num} - \sigma_\text{an}\right|$ to the angular momentum $\left| \sigma_{\text{disk}} \right|$ contained in the disk. This relative comparison is visible on the right panel of Fig. \ref{conv_num_fig_all}. A value close to $1$ means that $\Delta \sigma$ is negligible compared to the momentum in the disk. A value of $0$ or lower means that $\left| \Delta \sigma \right|$ is equal or higher than $\left| \sigma_{\text{disk}} \right|$. In this latter case, we cannot exclude the worst scenario where all the numerical errors would be concentrated at the same location, resulting in the formation of a disk which is then a numerical artifact. The right panel of Fig. \ref{conv_num_fig_all} shows the results until the disk starts to fragment, as our disk isolation algorithm gives correct results only when a single disk is present in the simulation box. These results show that the simulation with $\varepsilon = 50\%$ is the best in term of trust level. For the other levels of perturbation, the trust level is lower, but as this relative difference remains higher than $0.1$, even the worst scenario cannot explain the amount of angular momentum present in the disk in term of numerical errors.

\noindent
\end{appendix}
%============================================================================

\end{document}